\documentstyle[11pt,a4,cite,epsf]{article} 
\newcommand{\lesssim}{
{\ \lower-1.2pt\vbox{\hbox{\rlap{$<$}\lower5pt\vbox{\hbox{$\sim$}}}}\ } 
}
\newcommand{\gtrsim}{
{\ \lower-1.2pt\vbox{\hbox{\rlap{$>$}\lower5pt\vbox{\hbox{$\sim$}}}}\ } 
} 

\newcommand{\be}{\begin{equation}}
\newcommand{\ee}{\end{equation}}
\newcommand{\bea}{\begin{eqnarray}}
\newcommand{\eea}{\end{eqnarray}}
\newcommand{\noi}{\noindent}
\newcommand{\nn}{\nonumber}

\newcommand{\cO}{{\cal O}}

\newcommand{\cA}{{\cal A}}

\newcommand{\cD}{{\cal D}}
\newcommand{\cT}{{\cal T}}
\newcommand{\cP}{{\cal P}}
\newcommand{\cK}{{\cal K}}
\newcommand{\CF}{C_{\rm F}}

\newcommand{\alphak}{\alpha_{\mbox{\rm eff}}(k_{E}^2)}
\newcommand{\alpham}{\alpha(\mu^2)}
\newcommand{\alphamu}{\alpha(\mu^2)}
\newcommand{\alphaq}{\alpha(Q^2)}
\newcommand{\al}{\alpha}
\newcommand{\la}{\lambda}
\newcommand{\m}{\mu}
\newcommand{\n}{\nu}

\newcommand{\tr}{\mbox{\rm tr}}
\newcommand{\ks}{\not \! k}
\newcommand{\qs}{\not \! q}

\newcommand{\eff}{\mbox{\rm eff}}
\newcommand{\UV}{\mbox{\rm \small UV}}

\newcommand{\ra}{\rightarrow}

\newcommand{\PRD}[3]{Phys.\ Rev.\ {\bf D{#1}} (19{#2}) {#3}}

\def\theequation{\arabic{section}.\arabic{equation}}
 
\begin{document}
 
\begin{titlepage}
\begin{flushright} CPT-97/P.3451\\ UAB-FT-408\\ \today
\end{flushright}
\vspace*{2cm}
\begin{center} {\Large \bf Low-Energy QCD and Ultraviolet Renormalons
}\\[1.5cm] {\large {\bf Santiago
Peris}$^{a}$\footnote{Work partially supported by research project
CICYT-AEN95-0882.} and  {\bf Eduardo de Rafael}$^b$}\\[1cm]

$^a$ Grup de F\'{\i}sica Te\`orica and IFAE\\ Universitat Aut\`onoma
de Barcelona, 08193 Barcelona, Spain.\\[0.5cm] and\\[0.5cm]
$^b$  Centre  de Physique Th\'eorique\\
       CNRS-Luminy, Case 907\\
    F-13288 Marseille Cedex 9, France\\
\end{center}

\vspace*{1.0cm}
\begin{abstract}

We discuss the contribution of ultraviolet (UV) renormalons in QCD to
two--point functions of quark current operators. This explicitly 
includes effects due to the exchange of one renormalon chain as well as two
chains. It is shown that, when the external euclidean momentum of the
two--point functions becomes smaller than the  scale $\Lambda_L$
associated with the Landau singularity of the QCD one--loop running coupling
constant, the positions of the UV renormalons in the Borel plane 
become true singularities in the integration range 
of the Borel transform. This introduces ambiguities in the
evaluation of the corresponding two--point functions. The ambiguities 
associated with
 the leading UV renormalon singularity are of the same type as
the contribution due to the inclusion of dimension $d=6$ local
operators in a low--energy effective Lagrangian valid at scales smaller
than $\Lambda_L$. We then discuss the inclusion of an infinite number of
renormalon chains and argue that the previous ambiguity 
hints at a plausible approximation scheme for low--energy QCD, 
resulting in an effective Lagrangian similar to the one of the  
extended Nambu--Jona-Lasinio (ENJL) model of QCD at large $N_c$.         

\end{abstract}

\vfill

 
\end{titlepage}

\section{Introduction.}
\label{sec:Int}

The origin of renormalons in quantum field theory goes back to early work by 't
Hooft~\cite{tH78}, Lautrup~\cite{La77} and Parisi~\cite{Pa77}.  They made the
observation that in renormalizable theories like for example  quantum
electrodynamics (QED), there exists a class of Feynman diagrams which give rise
to a characteristic pattern in the coefficients of the large--order  terms in
the  perturbation series; these coefficients have an  
$n!$--growth with the same sign indicating that the corresponding series
are not Borel summable.~\footnote{For a comprehensive review of the subject and
a collection of  articles previous to 1990 see ref.~\cite{LeGZJ90}.} 
In QED this
growth originates in the large momentum integration region of virtual photons
dressed with vacuum polarization insertions which leads to singularities in the
associated Borel plane; the so--called  UV
renormalons~\cite{tH78,La77,Pa78,Co81,BS96}.  In quantum chromodynamics (QCD),
it is the low momentum integration region 
of virtual gluons dressed with running
couplings what is the source of the non integrable singularities in the
Borel plane; the so--called IR renormalons~\cite{tH78,Pa79,Mu85}. 
The appearance
of these singularities is perhaps not surprising since in the extreme kinematic
regimes in question it is not expected that these theories 
are well described by
a simple perturbative expansion in the coupling constant.

There are two good reasons to concentrate on the class of Feynman graphs which
are at the origin of the renormalon singularities. One is the fact that
in massless theories, like QCD with massless quarks, the breaking of conformal
invariance is encoded in the $\beta$--function. If one wants to find hints from
perturbation theory on the possible origin of {\it scales} it seems appropriate
to focus one's attention on the summation of the infinite subset of graphs
associated with the renormalization of the coupling constant. The other reason
is the opportunity that is offered to explore issues which have to do with
the analyticity (or rather, lack of it) in the coupling constant.    

The study of renormalon properties in gauge theories is at present an active
field of research. The r\^ole of IR renormalons in the operator product
expansion of two--point functions and their relationship with non--perturbative
inverse power corrections has been extensively discussed in the
literature~\cite{Da82,Da84,NSVZ84,NSVZ85,Mu85,Da86,Mu93,Sj95}. This and further
discussions which originate in the work of ref.~\cite{BY92}, has led to a new
point of view~\cite{Za92} concerning renormalons in QCD. 
The focus is now on the
possibility that their systematic study in a given hadronic process might
suggest generic non--perturbative effects of a universal nature. The
basic idea is that genuinely non--perturbative effects ought to cure
[Aany disease which appears when perturbation theory results are
analytically continued~\cite{AZ96}. This applies to the case of IR renormalons
and, as recently suggested by Vainshtein and Zakharov(V-Z)~\cite{VZ94a,VZ94b},
perhaps to the   much less explored r\^ole of UV renormalons in QCD as well.

\vspace*{0.5cm}

In QED (or for that matter in any renormalizable theory which is not
asymptotically free) the UV
renormalons cause a real obstruction to defining the  theory in the ultraviolet
{}from a resummation of the perturbative series. In earlier work by
Parisi on $\phi^4$ theories~\cite{Pa78} it was argued that it ought to be possible to mimic the
contribution from large momenta (namely UV renormalons) by means of insertions
of local dimension six and higher operators~\cite{BS96}. Specifically, local
here means that all the physics at momentum larger than a certain scale
$k$  is encoded in higher dimensional composite operators that are local on a
scale of
$1/k$. In other words, these composite operators 
are products of fields evaluated
at the same point but suppressed by inverse 
powers of the cut--off momentum $k$.
On these grounds, one would expect that by adding carefully--adjusted
higher--dimensional operators to the initial Lagrangian one ought to be able to
remove all the UV renormalons from QED to all orders in the coupling
constant~\cite{Gru95}. The intuitive reason 
is that QED can always be thought of
as the low--energy  limit of  a larger asymptotically free theory (like e.g.
$SU(5)$) for which, in principle, UV renormalons are not a problem. Of course
this means adding an infinite tower of higher dimensional operators which will
make QED an effective nonrenormalizable theory, in agreement with its alleged
triviality~\cite{ET96}. Notice that the previous argument is entirely
perturbative. In other words, the higher dimensional operators in question are
still considered as irrelevant. If, on the other hand, a few of them turned out
to be relevant, it might be possible to truncate the list of operators to
just the relevant ones. This is actually what happens in the quenched ladder
approximation of QED~\cite{Ba86} where four--fermion operators of the
Nambu--Jona-Lasinio type~\cite{NJL61} (NJL)  turn out to become relevant
operators  and consequently are kept in the Lagrangian on an equal footing as
the four dimensional ones. 

In this connection we wish to point out that there has also been an ongoing
struggle in the lattice community in trying to clarify if an abelian theory in
the strong coupling regime has anything to do with the dynamics of models
\`{a} la  Nambu--Jona-Lasinio.  Although this point is still unclear~\cite{Az},
there seems to be some evidence that four--fermion operators may play an
indispensable r\^{o}le for understanding the QED non--perturbative (i.e.
ultraviolet) dynamics.

\vspace*{0.5cm}

In QCD, and provided that the external momenta of a given Green's function are
not in regions of exceptional momenta~\cite{Sy}, the UV renormalons  are not an
obstruction to integrate virtual Euclidean momenta. Therefore one does not
expect them to be at the origin of fundamental ambiguities. In their recent
work~\cite{VZ94a,VZ94b}, V-Z have found however that the contribution to the
Adler function from the leading UV renormalon coming from the exchange of two
chains of vacuum polarization self--energy 
loops in an Abelian--like model  was,
contrary to naive expectations, dominant over the contribution coming from a
single chain. Furthermore, they also argued that the contribution from more and
more chains should be equally important 
and that, consequently, the actual value
of the  residue of the leading UV renormalon was ill defined. They then 
concluded 
that, at the phenomenological level, 
this could be taken as an indication that a new type of ambiguity may appear 
, which 
in the particular case of a two--point function with Euclidean momentum
$Q^2$, 
shows up as possible $1/Q^2$--like contributions.
 
Interestingly enough, the leading UV behaviour found by V-Z was shown to
originate in  the contribution from the insertion of $d=6$ four--fermion
operators. The appearance of these  four--fermion operators, much like
those of the Nambu--Jona-Lasinio model~\cite{NJL61}, is rather intriguing. 
It is known that there are extensions of this model 
which, when taken as models of QCD
in the large--$N_c$ limit at intermediate scales 
$\lesssim\cO(1\mbox{\rm GeV})$, are rather successful~\cite{BBdR,Bij} in
predicting low--energy physics (like for instance the $L_i$ coupling constants
of the Lagrangian of Gasser and Leutwyler~\cite{GaL}.) However, the possible
connection between these models and QCD has remained so far a mystery.

\vspace*{0.5cm}

There are some alternative routes to NJL--type models which have also been
suggested to describe low--energy QCD~\cite{MaS94,Kneur96}. Although not yet
comparable at the level of phenomenological success, they are  nevertheless
interesting in the sense that they represent relatively small departures from
perturbative  QCD and, hence, they ensure that  at least in {\it some
limit} they are likely to be related to QCD. Needless to say, the tough problem
they face is to show that the departures from perturbative QCD are big enough
and in the right direction to explain the observed phenomenology at low
energies.

A very popular model for instance is
that of a ``freezing'' coupling constant\cite{MaS94}. Although certainly
economic and rather successful, this approach (at least in its
most na\"{\i}ve version) runs into conflict with spontaneous chiral
symmetry breaking as we shall show with a particular example in the Appendix B. 

Another alternative approach to low--energy
QCD is the one which has recently appeared in refs.~\cite{Kneur96}. These
authors discuss an interesting attempt at describing spontaneous chiral
symmetry breaking in QCD  using a variational approach on resummed perturbation
theory. We are curious to know how this variational
approach could be related to the properties of large orders in the coupling
constant, i.e. to renormalons, and hence to some of the results that will be
obtained in this paper.

\vspace*{0.5cm}

The aim of this article is to study more 
closely and within a specific class of QCD
diagrams the interplay between the insertion of four--fermion operators and the
leading UV renormalon contributions; as well as the possible impact of this
relationship on bridging the gap between QCD and the low--energy chiral 
effective Lagrangian.  This we do by 
explicitly studying UV renormalon effects in
two--point functions of colour--singlet vector currents and of  colour--singlet
pseudoscalar currents. The currents which we consider are light quark currents
of the flavour
$SU(3)_{L}\times SU(3)_{R}$ group. Their associated two--point functions
correspond to physical observables in hadron physics. We are not so much
interested in UV renormalons as a source of possible
$1/Q^2$--like ambiguities in two--point function QCD sum rules, but rather in
their possible relevance to genuinely non--perturbative effects which they may
signal when the external Euclidean momentum $Q^2$ in a two--point function is
taken much smaller than the characteristic QCD scale. With the restriction to a
one--loop $\beta$ function this means
$Q^2 < \Lambda_L^2$, where $\Lambda_L^2$ is the Landau pole. Although this is
not the conventional situation (wherein $Q^2$ is always assumed to be very
large), we think that taking $Q^2$ small is unavoidable if one ever
wants to make any contact with a low--energy effective Lagrangian for QCD, as
this is the range of momentum which the effective Lagrangian is supposed to
describe.   We find that the ambiguities generated by the leading UV renormalon
in this regime of low
$Q^2$ hint at the existence of non--perturbative effective four--fermion local
operators to describe low--energy physics, much the same as the study of the
ambiguities generated by the IR renormalons hint at the existence of vacuum
condensates of composite operators in the OPE. 
Moreover, these non--perturbative
contributions turn out to be exactly like those of the extended NJL model
mentioned earlier, with the four--fermion operators normalized by an energy
scale which is momentum--independent, i.e. a constant. 
This constant which plays
the r\^{o}le of a cutoff in the Euclidean momentum integrals appears to be
related to the characteristic QCD scale and 
it seems natural to identify it with
$\Lambda_{\chi}$, the chiral symmetry breaking scale~\cite{GeM}. At the
level of a one--loop
$\beta$ function this identification implies that $\Lambda_{\chi}\simeq 
\Lambda_L$.

Most of the technical part of the paper has to do with the calculation of the
contribution to the residue of the leading UV renormalon coming from the {\it
effective charge} exchange of one and two--chains of gluon self--energy--like
graphs in these two--point functions. The concept of a QCD {\it effective
charge} at the one loop level is reviewed in Section~\ref{sec:Eff}, where we
also define the QCD ``amputated'' action which we use as a calculational
framework. For the sake of  simplicity we work in the chiral limit where the
light $u$, $d$, $s$  quark masses are neglected. In this limit the two--point
functions of flavour non--singlet axial--vector
 currents and scalar currents are
trivially related to those of vector currents and pseudoscalar currents
respectively; and it is therefore sufficient to study the two types of
two--point functions which we do.  Since, eventually, we are interested in a
comparison with an ``all--orders'' analysis of QCD in the large--$N_c$ limit,
we have kept track explicitly of the
$N_c$ factors which appear at the various stages of the calculations. The
calculations reported in Section~\ref{sec:UVTPF} are made in two ways. 
One method
uses the Gegenbauer expansion technique of conventional Feynman diagrams; the
other makes use of the operator product expansion technique following
refs.~\cite{VZ94a,VZ94b}. We find that the dominant r\^{o}le played by the
dimension--six four--fermion operators is indeed universal and comes about
in the
same way in the two channels which we have studied. The specific discussion of
the calculations at low $Q^2$ values is done in Section~\ref{sec:UVLow}, and it
is followed by Section~\ref{sec:ConOut}, which is dedicated to 
the conclusions and outlook.

\setcounter{equation}{0}

\section{The QCD Effective Charge and Renormalon Calculus.}
\label{sec:Eff}

In QED, the infinite subset of radiative corrections summed in the 
Dyson
series generated by the one--particle--irreducible vacuum polarization
self--energy function $\Pi_{R}(k^{2})$ defines an {\it effective 
charge} which
is universal, gauge--, and scheme--independent to all orders in 
perturbation
theory:
\be\label{QEDalphaeff}
\alpha_{\mbox{\small{\rm eff}}}(k^{2}) =
\frac{e_{R}^{2}}{4\pi}\,\frac{1}{1 + \Pi_{R}(k^{2})} =
\frac{e^{2}}{4\pi}\,\frac{1}{1 + \Pi(k^{2})}\,,
\ee 
where $e$ and $\Pi(k^{2})$ denote  bare 
quantities.
The extension of a similar {\it effective charge} concept to QCD  is of
fundamental importance for renormalon calculus, if one wants to identify
unambiguously the infinite subset of {\it gluon self--energy--like} radiative
corrections that one is summing in the replacement of a gluon propagator by a
so-called 
``renormalon chain''. Recently, there has been substantial progress in this
direction. The theoretical framework which has enabled this progress is the
so--called  pinch
technique~\cite{Co82,PTall,Wa96,PdeRW}. The pinch technique is a 
well--defined
algorithm for the rearrangement of conventional gauge--dependent 
one--loop
$n$--point functions to construct individually gauge--independent 
``one--loop'' $n$--point--like functions. This rearrangement of perturbation 
theory
is based on a systematic use of the tree level Ward identities of the
theory to cancel in Feynman amplitudes all factors of longitudinal
four--momentum associated with gauge fields propagating in loops. In 
the
case of QCD, the resulting {\it effective charge} has the form~\cite{Wa96}
\be
\label{eq:effchg}
\alpha_{\mbox{\small{\rm eff}}}(k_{E}^2)=\frac{g_{R}^{2}}{4\pi}
\frac{1}{1-\hat{\Pi}_{R}(k^2)}\,,
\ee
where $k_{E}^2\equiv -k^2\ge 0$ for $k^2$--spacelike, and 
$\hat{\Pi}_{R}(k^2)$
is a gauge invariant {\it gluon self--energy--like} two--point function, 
which
in the $\overline{MS}$--scheme in particular, and at the one loop level
is given by the expression
\be
\hat{\Pi}_{\overline{MS}}(k^2)=
\frac{g_{\overline{MS}}^2}{4\pi^2}\left\{\left(\frac{-11}{6}N_c +
\frac{1}{3}n_f
\right)\frac{1}{2}\log\frac{-k^2}{\mu^2}+
\frac{67}{36}N_c-\frac{5}{18}n_f\right\}\,.
\ee
The coefficient $\frac{-11}{6}N_c + \frac{1}{3}n_f$ of the logarithmic term
in this equation is precisely the first coefficient $\beta_{1}$ of the QCD
$\beta$--function. The {\it effective charge} encodes therefore the physics
of the $\beta$--function; in particular the scale breaking property. Since,
eventually, we are interested in the appearance of physical scales in QCD, it
seems natural to focus our attention on the properties of those diagrams of
perturbation theory generated by the insertion of {\it effective
charge}--exchanges. Of course, we do that at the simplest level of keeping
only the one loop dependence of the $\beta$--function. As far as one is only 
interested in general qualitative features this should not be a
serious limitation.

With the concept of an {\it effective charge} in hand it is
possible to define a framework to do QCD renormalon calculations in the sector
of light quark flavours. It is this framework which we next describe.

\vspace*{0.5cm} 

Let $\Gamma(v,a,s,p)$ be the full QCD generating functional of the
Green's functions of quark currents in the presence of external vector 
$v$,
axial--vector $a$, scalar $s$, and pseudoscalar $p$ matrix field sources:
\bea
\label{eq:GFQC} e^{i\Gamma(v,a,s,p)}
 & = & \frac{1}{\cal Z}\int\cD G_{\mu}\exp\left(-i\int dx\,\frac{1}{4}
G_{\mu\nu}^{(a)}(x)G^{(a)\mu\nu}(x)\right)\times \nn \\
 & & \int\cD\bar{q}\cD q\exp\left( i\int d^{4}x\,\bar{q}(x)i\not\!\!D
q(x)\right)\,,
\eea
where $\bar{q}=(\bar{u},\bar{d},\bar{s})$, and $\not\!\!D$ 
denotes the QCD Dirac
operator in the presence of the external sources
\be
\label{eq:Dirac}
\not\!\!D\equiv 
\gamma^{\mu}[\partial_{\mu}+ig_{s}G_{\mu}(x)]-i\gamma^{\mu}
[v_{\mu}(x)+\gamma_{5}a_{\mu}(x)]+i[s(x)-i\gamma_{5}p(x)]\,;
\ee
$G_{\mu}$ is the gluon gauge field colour matrix ($a=1,2,3,\dots 8$)
\be G_{\mu}(x)\equiv\sum_{a}\frac{\lambda^{(a)}}{2}G_{\mu}^{(a)}(x)\,;
\ee
and
\be 
G_{\mu \nu}^{(a)}(x)=\partial_\mu G_{\nu}^{(a)}-\partial_\nu G_{\mu}^{(a)}
-g_s f_{abc}G_{\mu}^{(b)}G_{\nu}^{(c)}\,,
\ee
the eight gluon field strength tensor components. The factor $\cal Z$ in
eq.~(\ref{eq:GFQC}) is such that $\Gamma(0,0,0,0)=1$.  

The QCD contributions which we shall consider in the renormalon
calculations presented below are the ones generated by an ``amputated''
generating functional $e^{i\tilde{\Gamma}(v,a,s,p)}$ defined as follows: 
\bea
\label{eq:appr} 
\lefteqn{e^{i\tilde{\Gamma}(v,a,s,p)} =} \nn \\
 &   &  
\frac{1}{\cal
Z}\int {\cal D}\bar{q}{\cal D} q \exp
\left\{i\int d^4x\, \bar{q}(x)\left[i\gamma^{\mu}
\left(\partial_{\mu}-i(v_{\mu}+\gamma_{5}a_{\mu})\right)
+i(s-i\gamma_{5}p)\right]q(x)\right\} \nn \\ & &  
\times\exp\left\{-g_{R}^{2}\sum_{a,b}\int d^4x\, d^4y\;
\bar{q}(x)\gamma^{\mu}\frac{\lambda^{(a)}}{2}q(x)\;
i\Delta^{ab}_{R\mu\nu}(x-y)\; 
\bar{q}(y)\gamma^{\nu}\frac{\lambda^{(b)}}{2}q(y)\right\}\,, 
\eea
with
$i\Delta^{ab}_{R\mu\nu}(x-y)$ the Fourier transform of the renormalized 
{\it gluon propagator--like} function:
\be
\label{eq:GBP} i\Delta^{ab}_{R\mu\nu}(x-y)=
\int\frac{d^{4}k}{(2\pi)^4}\, e^{-ik\cdot (x-y)} 
\frac{i\delta^{ab}}{k^2
+i\epsilon}
\left\{\left(-g_{\mu\nu}+\frac{k_{\mu}k_{\nu}}{k^2}\right) d_{R}
(k^2)-\xi\frac{k_{\mu}k_{\nu}}{k^2}\right\}\,,
\ee
and $d_{R}(k^2)$ the {\it effective charge} defined in
eq.~(\ref{eq:effchg}) i.e.
\be
\label{eq:effch}
g_{R}^{2}d_{R} (k^2)=\frac{g_{R}^{2}}{1-\hat{\Pi}_{R}(k^2)}\equiv
4\pi\alpha_{\mbox{\small{\rm eff}}}(k_{E}^2)\,.
\ee
The function
$i\Delta^{ab}_{R\mu\nu}(x-y)$ results from having integrated out those
gluonic interactions which, order by order in perturbation theory, 
contribute
to the QCD {\it effective charge} as successive powers of {\it gluon
self--energy--like} two--point functions evaluated at the one loop level. In
terms of a characteristic scale, like e.g. the
$\Lambda_{\overline{MS}}$ scale of the
$\overline{MS}$--scheme, we have
\be
\alpha_{\mbox{\small{\rm eff}}}(k_{E}^2)=\frac{1}{\frac{-\beta_{1}}{2\pi}
\log\frac{k_E^2}{c^2\Lambda_{\overline{MS}}^2}}\,,
\ee
where
\be
\beta_{1}=\frac{-11}{6}N_c +\frac{1}{3}n_f\qquad
\mbox{\rm and~\cite{Co82,PTall,Wa96,PdeRW}} \qquad
c^2=\exp\left\{\frac{67N_c -10n_f}{33N_c -6n_f}\right\}\,.
\ee
The numerical value of the constant $c$ in different relevant 
limits for $N_c$ 
and
$n_f$ does not change much: 
$c(N_c=3,~n_f=3) =  2.87\,, c(N_c\rightarrow\infty) = 2.76\,,
c(n_f\rightarrow\infty) = 2.30\,.$
We shall often refer to the scale
\be
\label{eq:lambda}
\Lambda_L=c\times \Lambda_{\overline{MS}}\,,
\ee
as the Landau pole. Notice that, 
with $\Lambda_{\overline{MS}} \sim 300-400$ MeV, as determined
phenomenologically, $\Lambda_L$ turns out to be numerically 
of the same size as the chiral symmetry breaking scale: 
$\Lambda_{L}\simeq  
\Lambda_{\chi}\simeq 1\mbox{\rm GeV}$ . Although c and
$\Lambda_{\overline{MS}}$ are scheme dependent, the combination 
$\Lambda_L$
is scheme independent. For simplicity we shall work in a subtraction 
scheme
in which the corresponding c is unity, or equivalently, wherein 
$g_R(\mu)=g_{\overline{MS}}(\mu/c)$; and from now on we shall drop the 
subscript ``R" from $g_R(\mu)$.

We consider that the framework described above is a net 
improvement
with respect to previous approaches to renormalon calculus in QCD. The
existence of an {\it effective charge} in QCD, with properties analogous to 
those of the QED
{\it effective charge}, as recently emphasized in refs.~\cite{Wa96}
and~\cite{PdeRW}, provides the basic feature that allows us 
to select a minimal class
of well defined  contributions to QCD renormalons. In the so called na\"{\i}ve
non--abelianization procedure (NNA)~\cite{Br93},  it is
suggested to perform first QED--like calculations ---ignoring the  non--abelian
gluonic interactions--- and replace in the results thus obtained the number
of light quark flavours, $n_{f}$, coming from QED--like vacuum polarization 
insertions, by
\be 
n_{f}\ra n_{f}-\frac{11}{2}N_{c}\,.
\ee
Another procedure often adopted is to consider abelian--like 
gluonic
interactions dressed with the asymptotic running coupling at the scale 
of the
virtual gluonic Euclidean momentum, and then advocate a  ``large
$\beta_{1}$ expansion''.~\footnote{See e.g. ref.~\cite{LTM95} and 
references
therein.} None of these prescriptions, included the ``amputated'' generating
functional in (\ref{eq:appr}) which we are proposing, has been justified as
yet in terms  of a well defined  approximation within QCD itself. The merit
however of the ``amputated'' effective action in (\ref{eq:appr}) is that it
selects a minimum set of  a well defined class of QCD
contributions; and in principle it could be improved by considering 
higher
loop contributions to the {\it gluon self--energy--like} function
$\hat{\Pi}_{R}$ which governs the {\it gluon
propagator--like} function
$i\Delta^{ab}_{R\mu\nu}(x-y)$; as well as by taking into consideration 
more and more non--local interaction (three--point--like, 
four--point--like, ...) terms.

\vspace*{0.5cm}   

As an illustrative example, we dedicate the rest of this Section to
the calculation of the contribution to the Adler function induced by the
renormalon effects which result from the exchange of an 
{\it effective charge}
chain in the ``amputated'' generating functional 
$\tilde{\Gamma} (v,a,s,p)$ in
(\ref{eq:appr}). 
With 
\be  V^{\mu}=\frac{1}{2} \ 
(:\bar{u}\gamma^{\mu}u:-:\bar{d}\gamma^{\mu}d:) 
\ee
the vector--isovector quark current, and
\bea
\label{eq:pi1}
\Pi^{\mu\nu}(q) & = &  i\int\, d^4 x e^{iq\cdot x}\langle 0\mid
 \mbox{\rm T}\left\{V^{\mu} (x)V^{\nu}(0)\right\}\mid 0\rangle \nn \\
 & = & -\left(g^{\mu\nu} q^2-q^{\mu}q^{\nu}\right)\Pi(q^2)
\eea
the associated correlation function, the Adler function is defined as 
the
logarithmic derivative of $\Pi(q^2)$: ($Q^2\equiv -q^2$, with $Q^2>0$ for
$q^2$--spacelike)
\be 
\label{eq:Adler} {\cal A}(Q^2) \equiv -Q^2\frac{\partial \Pi}{\partial 
Q^2}\,.
\ee
In perturbative QCD calculations, $Q^2$ is supposed to be larger
than the characteristic QCD scale, which in our case is $\Lambda_L^2$ in
eq.~(\ref{eq:lambda}). In terms of Feynman--like diagrams, the calculation in
question can be represented by diagrams like the one in Fig.~1:

\vspace{5mm}

\centerline{\epsfbox{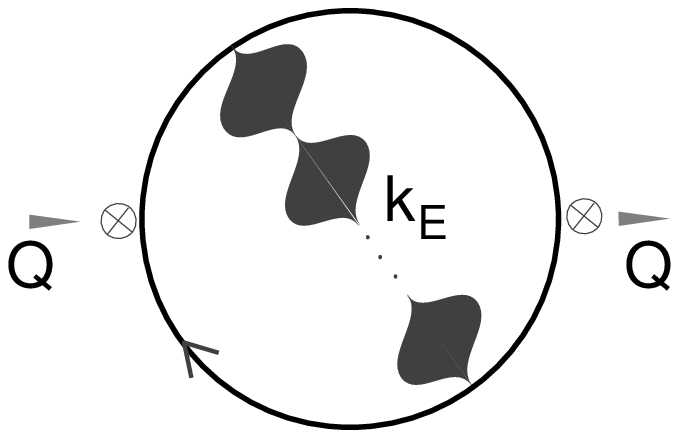}}
\vskip 1pc {{\bf Fig.~1} {\it One of the three Feynman--like diagrams with an
effective-charge chain which induce renormalon effects
in the Adler function.}} 
\vspace*{8mm} 

\noi The chain of bubbles in Fig.~1  corresponds
to the replacement of the ordinary free gluon propagator by the 
full {\it gluon propagator--like} function 
\be
\label{eq:gr}
\frac{-i\left(g_{\mu\nu}-(1-\xi)
\frac{k_{\mu}k_{\nu}}{k^2}\right)}{k^2  + i\epsilon}(-ig_{s})^2
\Rightarrow
-i\left(g_{\mu\nu}-\frac{k_{\mu}k_{\nu}}{k^2}\right)
\frac{4\pi\,\alpha_{\mbox{\small{\rm eff}}}(k_E^2)}{k_{E}^2-i\epsilon}
 - i\xi\frac{k_{\mu}k_{\nu}}{k^2}
\frac{(-ig_s)^2}{k^2 +i\epsilon}\,,
\ee
where $k_{E}$ denotes the Euclidean virtual momentum carried by the chain; 
and $\xi$ is the same covariant gauge
parameter as in eq.~(\ref{eq:GBP}). This replacement is in fact the net
effect, in momentum space, of the interaction term in 
eq.~(\ref{eq:appr})
when evaluated to its lowest non trivial order.

An interesting function to consider then is the functional derivative
\be
\frac{\delta\Pi(Q^2)}{\delta \alphak}\,.
\ee
Up to overall normalization factors, it corresponds to the forward
elastic scattering amplitude of an off--shell vector--isovector quark 
current
off an off--shell gluon evaluated at the one loop level as illustrated in
Fig.~2.

\vspace{5mm}

\centerline{\epsfbox{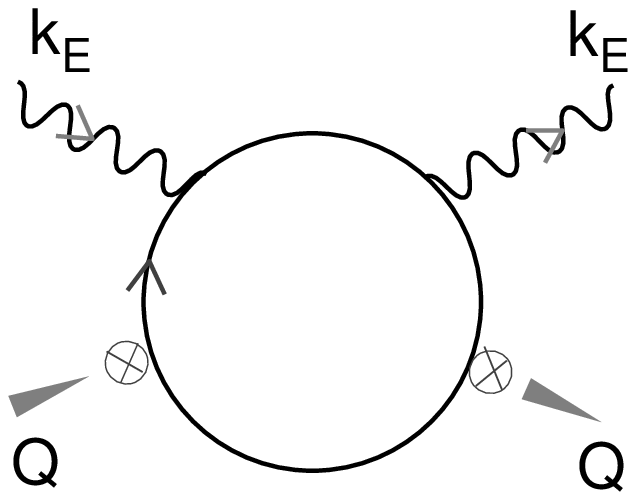}}
\vskip 1pc {{\bf Fig.~2} {\it One of the three Feynman diagrams
corresponding to the forward scattering amplitude of an off--shell
vector--isovector current, carrying Euclidean momentum $Q$, off an off--shell
gluon carrying Euclidean momentum $k_{E}$.}} 
\vspace*{8mm} 

\noi
Once this function is known, the full calculation of renormalon
effects induced by one {\it effective charge} chain exchange is simply given by
the integral 
\be
\label{eq:master}
\Pi(Q^2)= \int_{0}^{\infty}dk_{E}^2
\:\frac{\delta\Pi(Q^2)}{\delta \alphak} \:\alphak\,.
\ee
It turns out that the kernel given by the  functional derivative above
can be extracted from early papers on QED by Baker and 
Johnson~\cite{BJ69}.
It can be easily adapted to QCD by including the appropriate colour 
factors
and it has the following form $\left(\CF=\frac{N_c^2 -1}{2N_c}\right)$:
\be
\frac{\delta\Pi(Q^2)}{\delta 
\alphak}=\frac{N_c}{32\pi^3}\,\CF\,k_{E}^2\times
\left\{\begin{array}{ll}
\frac{1}{Q^4}\:\Xi\left(\frac{k_{E}^2}{Q^2}\right) & 
\mbox{for $k_{E}^2\le Q^2$}\,,
\\
\frac{1}{k_{E}^4}\:\Xi\left(\frac{Q^2}{k_{E}^2}\right) & 
\mbox{for $k_{E}^2\ge Q^2$}\,, 
\end{array}\right. 
\ee
clearly illustrating the UV $\rightleftharpoons$ IR symmetry of the kernel.
Notice that renormalization does not affect the calculation of this kernel.
At this level QCD with massless quarks still has its full conformal invariance
properties, which is reflected in the symmetries of the kernel above.  The
function
$\Xi(z)$ has the simple integral  representation~\cite{BJ69}
\be
\Xi(z)=1+\frac{4}{3}\int_{0}^{z}dy \left(1-\frac{1}{z}y\right)^2
\frac{1}{1+y}\log y\,.
\ee
In the interval $0\le z \le 1$ it is a monotonically decreasing function from
$\Xi(0)=1$ to
$\Xi(1)=\frac{14}{3}-\frac{4}{9}\pi^2 =0.2802...$, and it has the Taylor
expansion:
\be
\label{eq:xiex}
\Xi\left(z\right)  = 1+\frac{4}{3} z 
\left(\frac{1}{3}\log z -\frac{11}{18}\right) 
 -\frac{4}{3} z^2 \left(\frac{1}{12}
\log z -\frac{13}{144}\right)+\cdots\;. 
\ee

For the purpose of renormalon calculations, it is natural to separate the
integration region in eq.~(\ref{eq:master}) into an infrared--dominated region
where $0\le k_{E}^2\le Q^2$, and an ultraviolet--dominated region where $Q^2\le
k_{E}^2\le
\infty$.   The integral in the infrared region is in fact ill--defined because
the  function
$\alphak$ blows up  at the Landau pole where
$k_{E}^2=\Lambda_{L}^2$.~\footnote{This is in fact only true for a one--loop
$\beta$ function. In the case of a two--loop
$\beta$ function, the relevant 
pole is not at $\Lambda_{L}^2$~\cite{PdeR}, but the
physics is pretty much the same.} As a consequence of this, there is an  
obstruction as a
matter of principle to reconstructing the full non--perturbative answer 
{}from
resummed perturbation theory. This intuitive argument can be refined
and made more systematic with the use of Borel resummation
techniques~\cite{LeGZJ90}. Let us discuss the most salient features of 
the
integral in eq.~(\ref{eq:master}):

\begin{itemize}
\item{\it The limit $\alphak=\alpha(\mu^2)$.}

In this limit we  find the well known perturbative  result corresponding to the
one  gluon
exchange contribution to the Adler function, namely 
\be
\label{eq:adler}
\cA (Q^2)=\frac{N_c}{16\pi^2}\frac{2}{3}\left(1+
\frac{3}{4}\,\CF\,\frac{\alpha(\mu^2)}{\pi}\right)\,.
\ee
Notice that in this limit the integral in the region from
$0\le k_{E}^2\le Q^2$ contributes a constant to $\Pi(Q^2)$ and  hence 
nothing
to the Adler function; i.e. it is renormalized away.  One--gluon 
exchange with
no {\it gluon self--energy--like} correction (no bubble) in it is not
sensitive to the infrared region. Only the first 
term $\Xi\left(\frac{Q^2}{k_{E}^2}\rightarrow 0\right)=1$  in the expansion of
the
$\Xi$ function in the UV region, where $k_{E}^2> Q^2$, contributes to 
the Adler function in this limit.

\item{\it Infrared renormalons.}

Once we start summing bubbles corresponding to the expansion of $\alphak$ in
powers of {\it gluon self--energy--like} insertions, we find that  the leading
infrared renormalon is induced by  the first term, 
$\Xi\left(\frac{k_{E}^2}{Q^2}\rightarrow 0\right)=1$,   in  the 
expansion of the $\Xi$--function in the IR region. Making the change of
variables
\be 
w/2=- b_0
\alphaq \log k_E^2/Q^2\,,\qquad \mbox{\rm where}\qquad b_0\equiv
-\beta_1/2\pi  
\ee
in eq.~(\ref{eq:master}), one obtains 
\be
\Pi(Q^2)\vert_{\mbox{\small{\rm IR}}}
=\frac{N_c}{16\pi^2}\,\CF\,\frac{1}{2\pi b_0}
\int_{0}^{\infty}dw\, e^{-\frac{w}{b_0\alpha(\mu^2)}}\:\frac{1}{2-w}
\left(\frac{Q^2}{\mu^2}\right)^{-w}\,,
\ee
which leads to the following contribution to the Adler function
\be
\label{eq:irr}
\cA(Q^2)\vert_{\mbox{\small{\rm IR}}}
=\frac{N_c}{16\pi^2}\,C_{F}\,\frac{1}{2\pi b_0}
\int_{0}^{\infty}dw\, e^{-\frac{w}{b_0\alphamu}}\:\frac{w}{2-w}
\left(\frac{Q^2}{\mu^2}\right)^{-w}\,.
\ee

\noi This expression is already in the form of a Borel transform. An
expansion around $w=0$ would generate the characteristic $n!$ behaviour 
of the
perturbative expansion in $\alphamu$ which we mentioned in
the Introduction. As already discussed by other 
authors~\cite{Mu85,Za92,Be93}, we find that the leading IR renormalon
contribution to the Adler function appears as a pole in the Borel 
plane at $w=2$. There is no term in the IR expansion of the
$\Xi\left(\frac{k_{E}^2}{Q^2}\right)$ function which leads to a pole at
$w=1$. From the previous change of variables one also sees that low values 
of
$w$ are associated with momenta around the large scale $Q^2$, where
perturbation theory is expected to give a good description of the 
dynamics.
However, as $w$ goes up (and it has to go all the way up to infinity) one 
enters
deeper and deeper into the IR region, where perturbation theory must 
fail or
else QCD would not describe the spectrum of hadronic bound states that are 
observed.
The singularity at $w=2$ exhibits this in its crudest form.  More 
precisely,
this singularity implies an ambiguity in the  perturbative  evaluation 
of the
Adler function. Therefore, the analytic continuation in $w$ in
eq.~(\ref{eq:irr}), or equivalently the resummation of perturbation theory into
the effective charge $\alphak$ of eq.~(\ref{eq:master}), that  one has tried in
order to obtain the full solution has failed.
The ambiguity  is encoded in the unavoidable 
prescription to
skip the pole; and is of the form $\sim e^{-1/b_0 \alpha(Q)}$ :  
\be
\label{eq:ambiguity}
\delta\cA(Q^2)\vert_{\mbox{\small{\rm
IR}}}\sim\frac{N_c}{16\pi^2}\, \frac{\CF}{-\beta_1}
\left(\frac{\Lambda_L}{Q}\right)^4\,,
\ee
i.e., an ambiguity which has the same $\frac{1}{Q^4}$ pattern as the
gluon condensate contribution  that appears in the OPE evaluation of 
the
Adler function~\cite{NSVZ84,NSVZ85}. Since the Adler function must be an
unambiguous physical observable, there must exist another contribution that
cancels (\ref{eq:ambiguity}). 
In other words, (Borel) resummed
perturbation theory requires the presence of the gluon condensate 
(which
is also ambiguous for the same reason) to combine with the result of
eq.~(\ref{eq:irr}) and yield a final well--defined answer~\cite{Mu85}.  
This
is an example of how all--orders perturbation theory is capable of ``hinting''
at  non--perturbative
dynamics.

Following analogous steps, the insertion of higher powers in $k_{E}^2/Q^2$  
{}from the IR expansion of the
$\Xi$--function in the $0\le k_{E}^{2}\le Q^2$ 
integrand produces contributions to
higher singularities in the Borel plane corresponding to higher order
IR renormalons located at $w=3$, $w=4$, and so on. However, the higher power
terms of the $\Xi$--function  only generate partial contributions to the higher
order IR renormalons. For example, for the IR renormalon located at
$w=3$, there will also be further contributions originating in the 
three--point--like non--local term which 
will appear as a {\it correction} to the
``amputated'' generating functional in (\ref{eq:appr}). This new term will
generate diagrams like e.g. the one shown 
in Fig.~3 which we are not taking into
account here.

\vspace{5mm}

\centerline{\epsfbox{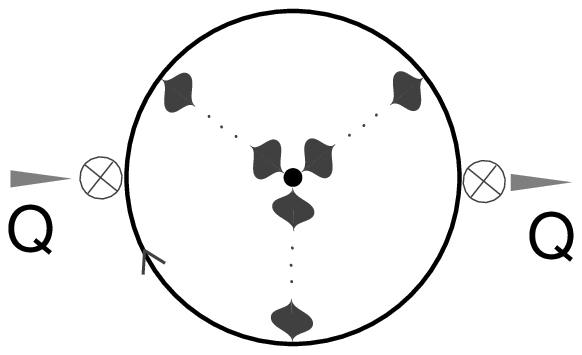}}
\vskip 1pc {{\bf Fig.~3} {\it Example of a Feynman diagram generated by
the three gluon--like vertex.}} 
\vspace*{8mm} 

\noi  The total contribution to
higher order IR renormalons is associated with  higher dimensional  condensates
with a larger power in
$1/Q^{2}$ than the one in (\ref{eq:ambiguity}), and in this sense 
they are non--leading.~\footnote{In fact, in the particular case of the IR
renormalon located at $w=3$, and with no extra hard $\alphaq$ correction, we
expect a cancellation of the various partial contributions to the  residue of
the pole at $w=3$, in accordance with the fact that there is no three gluon
condensate contribution in the OPE to the Adler function as proved in
refs.~\cite{DuSm81,HuMa82}.}   

\item{\it Ultraviolet renormalons.}

The integral in eq.~(\ref{eq:master}), in contrast to the behaviour in 
the
infrared region discussed above,  is well behaved in the ultraviolet 
region
where $Q^2\le k_{E}^2\le\infty$, provided that
$Q^2 >>
\Lambda_L^2$. Nonetheless it also has associated with it a Borel
representation, the only difference being that now the poles will occur 
at
negative values of the Borel variable $w$. As in the case of infrared
renormalons, the further away the pole is   from the origin
$w=0$ the smaller will be the $n$-th coefficient accompanying 
$\alphaq ^n$ in the perturbative series. This gives rise to a hierarchy 
of
ultraviolet renormalons. The leading one, within the exchange of one power of the
{\it effective charge}, is the one induced by  the 
${\cal
O}(z)$--term in the UV expansion ($k_E^2 > Q^2$) of the
$\Xi$--function in eq.~(\ref{eq:xiex}).  With the change of variables 
$w = b_0 \alphaq \log k_E^2/Q^2$ one obtains
\bea
\label{eq:UVRI}
\Pi(Q^2)\vert_{\mbox{\small{\rm UV}}} & = & -\   
\frac{N_c}{16\pi^2}\,\CF\,\frac{4}{9}\,
\frac{1}{2\pi b_0}
\int_{0}^{\infty}dw\,
e^{-\frac{w}{b_{0}\alpha(\mu^2)}}\,\left(\frac{\mu^2}{Q^2}\right)^{w} 
\nn \\
 & & \quad 
\left\{\frac{1}{(1+w)^2}+
\frac{11}{6}\frac{1}{1+w}\right\}\,,
\eea
where the leading
$\frac{1}{(1+w)^2}$ term is the one induced by the term 
$\frac{4}{3}\left(\frac{Q^2}{k_{E}^2}\frac{1}{3}\log\frac{Q^2}{k_{E}^2}
\right)$ in eq.~(\ref{eq:xiex}).  The corresponding  contribution to the 
Adler function is then
\bea
\label{eq:AdlerUV}
\cA (Q^2)\vert_{\mbox{\small{\rm UV}}} & = &
\frac{N_c}{16\pi^2}\,\CF
\,\frac{4}{9}\,\frac{1}{2\pi b_0}
\int_{0}^{\infty}dw\, e^{-\frac{w}{b_0\alphaq}} \nn \\
 & & \quad 
\left\{\frac{1}{(1+w)^2}+ \frac{5}{6} \frac{1}{(1+w)} - 
\frac{11}{6}\right\}\,,
\eea
where we have also scaled $\mu^2$ at $Q^2$. The result of the leading
term $\simeq\frac{1}{(1+w)^2}$ agrees with the one obtained in
refs.~\cite{Be93,VZ94a,VZ94b} using other methods.  One sees that the
singularity is at
$w=-1$, i.e. outside the integration range. Therefore, unlike the case 
of IR
renormalons, one cannot argue now that the singularity is an 
obstruction to do
the integral. This is also clear from the above change of variables 
since for
$0\le w \le \infty$ one never leaves the region of very large momentum 
where
perturbation theory is supposed to be a faithful description of the 
dynamics.
 
Contributions to higher--order UV renormalons can be obtained from the
insertion of  successive
higher power terms in the UV expansion of the $\Xi$--function in the
$k_{E}^2$ integrand in eq.~(\ref{eq:master}) for $Q^2\le k_{E}^2\le 
\infty$. The next Section is fully dedicated to the study of UV renormalons in
two--point functions. 

\end{itemize}

\newpage
\setcounter{equation}{0}

\section{UV Renormalon Contributions to Two--Point Functions.}
\label{sec:UVTPF}
We shall discuss in this Section the calculations of the contributions to
two--point functions of quark current operators induced by the leading
UV renormalon generated by the exchange  of one and two QCD {\it effective
charge} chains. The two--point functions which we shall consider are the Adler
function already discussed in the previous Section, and the two--point
function associated with the divergence of the axial--current:
\be
\label{eq:axial}
\partial^{\mu}A_{\mu}(x)\equiv (m_{d}+m_{u})
\ :\bar{d}(x)i\gamma_{5}u(x):\,,
\ee
where the overall quark mass factor will only be kept so as to make the
pseudoscalar current density renormalization invariant. All the calculations
presented here have been made in the chiral limit where the masses of 
the light $u$, $d$, and $s$ quarks are set to zero. 
The two--point function associated with the divergence of
the axial current is
\be
\label{eq:psifive}
\Psi_{5}(q^2)=i\int\,d^4 x e^{iq\cdot x}
\langle 0\mid \mbox{\rm T}\left\{\partial^{\mu}A_{\mu}(x) 
\partial^{\nu}A^{\dagger}_{\nu}(0)\right\}\mid 0\rangle \,.
\ee 
The equivalent of the Adler function here is the second derivative of
$\Psi_{5}(q^2)$,
\be 
\label{eq:psifiveder} {\cal P}(Q^2)\equiv \frac{Q^{2}}{(m_u +
m_d)^2}\frac{\partial^{2}\Psi_{5}(q^2)}{(\partial Q^2)^2}\,.
\ee 
This second derivative is, like the Adler function in
eq.~(\ref{eq:Adler}), independent of external subtractions.

\vspace*{1cm}     
\subsection{The Vector Two--Point Function.}
\label{subsec:VTPF}

As mentioned in Section~\ref{sec:Int} we have carried out  
the UV renormalon calculations
using two different methods. One is the operator product
expansion (OPE) technique, which we discuss below, following rather closely
the QED work of Vainshtein and Zakharov~\cite{VZ94a,VZ94b}. The other method
uses ordinary Feynman diagrams directly and the Gegenbauer polynomial expansion
of Euclidean propagators. For the vector two--point function, we shall  
discuss this second method only 
in connection with the exchange
of two {\it effective charge} chains.

\subsubsection{\normalsize {\bf OPE Calculation: One Chain.}}
\label{subsubsec:OPEOC}

Let us start with the contribution from the UV renormalon generated by the
exchange of one {\it effective charge} chain, to which we shall refer for short
as the contribution  of a ``one--renormalon chain'' or even ``one--chain''. The
contribution of the ``one--renormalon chain'' to the external vector field
vacuum polarization tensor of eq.~(\ref{eq:pi1})  which we have discussed in
Section~\ref{sec:Eff} can also be written as follows:
\be
\label{eq:pivector}
\Pi^{\m \n}(q)\ v^\m v^\n = \frac{-i}{2} \int_{k_E^2 \ge Q^2}
\frac{d^4k}{(2\pi)^4}\ \frac{4 \pi \alphak}{k^2}\ \langle v\mid
\frac{\cT}{k^2} \mid v
\rangle\,,
\ee 
where the external vector field $v$ is the same as the one appearing in
the Dirac operator in eq.~(\ref{eq:Dirac})  and $\cT$ is the time--ordered
operator 
\be
\label{eq:t}
\cT \equiv i\ k^2 \int d^4x e^{ik\cdot x}\ \mbox{\rm T}\left\{:\bar
q(x)\gamma^\m
\frac{\la^a}{2}q(x):\ :\bar q(0)\gamma_\m
\frac{\la^a}{2}q(0):\right\}\,.
\ee 
Following \cite{VZ94a,VZ94b} we now perform  the
operator product expansion on the operator $\cT$ by using the Schwinger
background field technique, explained e.g. in ref.~\cite{OPE}. The first
nonvanishing contributions comes from the dimension--six operators:
\be
\label{eq:tone}
\cT = \frac{4}{3k^2}\ \bar q(0) \gamma^\m \left[-\
\frac{g_{\eff}(k_{E}^2)}{2N_c}
\ D^\al G_{\al \m}(0) + \CF \ D^\al F_{\al \m}(0)\right] q(0)\,,
\ee 
where $F_{\m \n}(x)$ is the
field strength tensor associated with the external vector source. With this
expression for the operator
$\cT$ one can easily compute the integral in  (\ref{eq:pivector}). [Recall
eq.~(\ref{eq:pi1}), and notice that at this order only the
$F_{\al
\m}$ term
 in (\ref{eq:tone}) contributes.] This results in the expression
\be
\label{eq:ten}
\Pi(q^2)= -\ \frac{N_c}{16\pi^2}\ \frac{2}{9}\  Q^2 \int_{Q^2}^\infty
\frac{dk^2_E}{k_E^4}\ \frac{\alphak}{\pi} \ \log(k^2_E/Q^2)
\,,
\ee 
where the momentum in the argument of the logarithm has been cut
off at the scale $k_{E}^2$ that appears in the OPE in
eq.~(\ref{eq:tone}). The steps of the calculation are illustrated in Fig.~4
below:

\vspace{5mm}

\centerline{\epsfbox{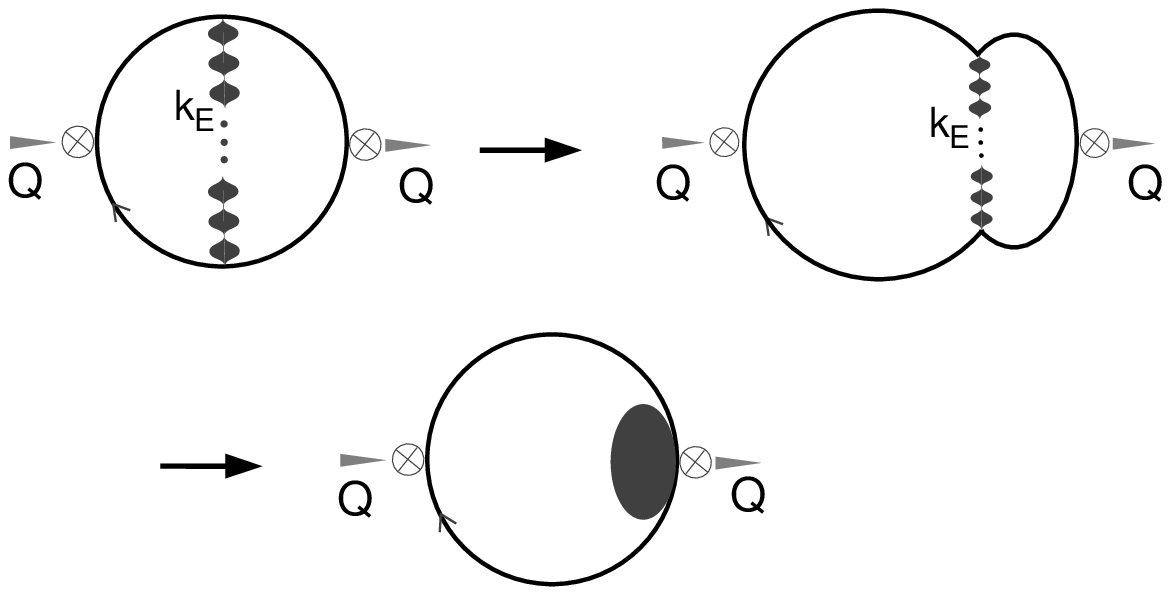}}
\vskip 1pc {{\bf Fig.~4} {\it The leading UV renormalon contribution from a one
effective charge exchange, evaluated with the operator product expansion
technique, is the one generated by the dimension six vertex like operators
simulated by the form factor in the lower figure.}} 
\vspace*{8mm} 

The UV renormalon contribution to the Adler function in (\ref{eq:Adler})
can now be readily obtained. Using the change of variables
$w=b_0
\alphaq \log(k_E^2/Q^2)$ results in
\be
\label{eq:Aonechain}
\cA(Q^2)\mid^{\mbox{\rm \small one \ chain}}_{\mbox{\rm \small UV}} = \
\frac{N_c}{16\pi^2}\,\CF\,\frac{4}{9}\,
\frac{1}{2\pi b_{0}}
\int_0^\infty \frac{dw}{(1+w)^2}\ e^{-\frac{w}{b_0 \alphaq}}
\ee 
which, if expanded in perturbation theory, yields
\be
\label{eq:Aonechainexpanded}
\cA(Q^2)\mid^{\mbox{\rm \small one \ chain}}_{\mbox{\rm \small UV}} \sim -\ 
\sum_{\mbox{\rm Large\ n}}\ 
\frac{N_c}{16\pi^2}\CF\, \frac{4}{9}\,\frac{1}{2\pi b_{0}}
\left(- b_0 \alphaq \right)^n n! \quad .
\ee
Notice that the result in (\ref{eq:Aonechain}) is in agreement with the one
we found in eq.~(\ref{eq:AdlerUV}) for the leading behaviour of the Borel
transform around the singularity
$w=-1$. Our result coincides also with the one obtained in
refs.~\cite{VZ94a,VZ94b} in the abelian limit where $\frac{\la^a}{2}
\rightarrow 1$.

\subsubsection{\normalsize{\bf OPE Calculation: Two Chains.}} 
\label{subsubsec:OPETC}

In an analogous way one can deal with the more complicated contributions
coming from ``two renormalon chains'', as illustrated in Fig.~5. Two different
new structures emerge from the box--like diagrams:

\bea
\cT^{\mbox{\rm \small Box}} & = &\cT_1^{\mbox{\rm \small Box}} 
+ \cT_2^{\mbox{\rm \small Box}}
 \nn \\
&= & -\
\frac{g^2_{\eff}(k_{E}^2)}{2k^2}\ \biggl[
\bar q(0)\gamma^\m \gamma^\la \gamma^\al \frac{\la^a}{2} \frac{\la^b}{2} q(0) 
\bar q(0)\gamma_\m \gamma_\la \gamma_\al \frac{\la^a}{2} \frac{\la^b}{2} q(0)
\nn \\ & & - \ 
\bar q(0)\gamma^\m \gamma^\la \gamma^\al \frac{\la^a}{2} \frac{\la^b}{2} q(0) 
\bar q(0)\gamma_\al \gamma_\la \gamma_\m \frac{\la^b}{2} \frac{\la^a}{2} q(0)
\biggr]\,.
\eea

\vspace{5mm}

\centerline{\epsfbox{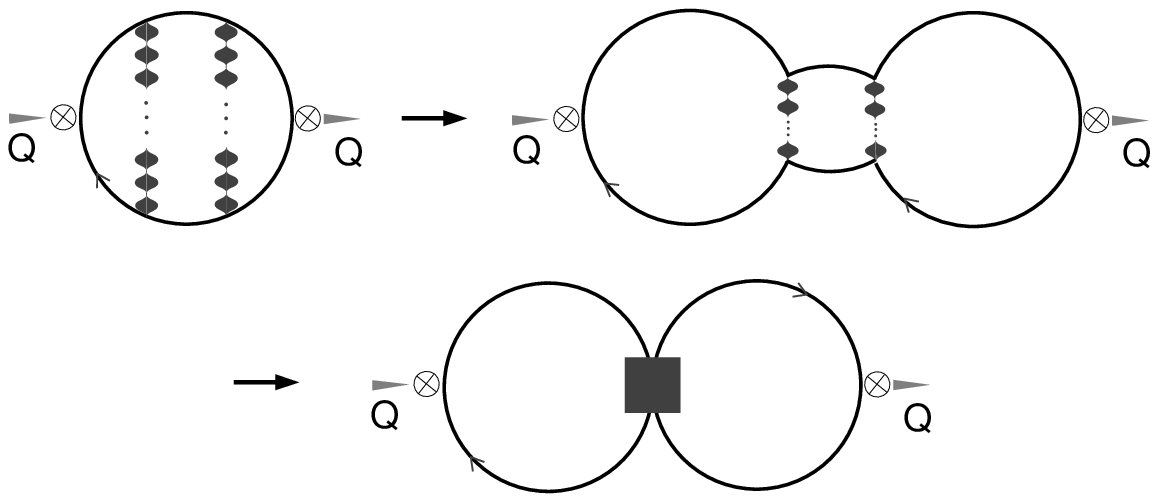}}
\vskip 1pc {{\bf Fig.~5} {\it The leading UV renormalon contribution from the
exchange of two effective charges. The  evaluation with the operator product
expansion technique, generates a dimension six four--fermion
operator simulated by the box form factor in the lower figure.}} 
\vspace*{8mm} 
 
\noi
Using the operator identity
\be
(\gamma^\m
\gamma^\la \gamma^\al) \bigotimes (\gamma_\al \gamma_\la \gamma_\m) = 10\
\gamma^\m \bigotimes \gamma_\m - 6\ \gamma^\m \gamma_5 \bigotimes \gamma_\m
\gamma_5\,,
\ee
one can immediately see that the abelian limit yields the result
\be
\label{eq:abelian}
\cT^{\mbox{\rm \small Box}}_{\mbox{\rm \small abelian}} = -\ \frac{6\
g^2_{\eff}(k_{E}^2)}{k^2}\
\bar q(0)\gamma^\m \gamma_5 q(0) \bar q(0)\gamma_\m \gamma_5 q(0)\,;
\ee 
i.e. there is a cancellation of the $\gamma^\m \bigotimes \gamma_\m $
structure. Equation (\ref{eq:abelian}) was also obtained by the authors of
refs.~\cite{VZ94a,VZ94b}. However, in the large--$N_c$ limit, the previous 
cancellation does not take place. Instead one sees that 
$\cT_2^{\mbox{\rm \small Box}}$ is proportional to $N_c$ 
whereas $\cT_1^{\mbox{\rm \small Box}}$ goes like
$1/N_c$, i.e. it is suppressed by two powers of $N_c$ relative to $\cT_2^
{\mbox{\rm \small Box}}$. Consequently, at large--$N_c$, 
\bea
\label{eq:BoxNc}
\cT^{\mbox{\rm \small Box}}& \simeq  & 
\cT_2^{\mbox{\rm \small Box}} \nn \\ & \simeq  &
\frac{N_c\ g_{\eff}^2(k_{E}^2)}{4k^2}\ \biggl\{ 10\ \bar q(0)
\gamma^\m \frac{\la^a}{2} q(0) \bar q(0) \gamma_\m \frac{\la^a}{2} q(0)
\nn \\  & & - 6\ \bar q(0)
\gamma^\m \gamma_5 \frac{\la^a}{2} q(0) 
\bar q(0) \gamma_\m \gamma_5 \frac{\la^a}{2} q(0) \biggr\}
\,.
\eea

Upon comparing with the one--renormalon chain contribution of
eq.~(\ref{eq:tone}) we see that the use of the equations of motion for the
gluon field in that expression  also produces  a four--fermion operator of the
form (\ref{eq:BoxNc})~\cite{VZ94a,VZ94b}~
\footnote{i.e. a penguin--like diagram.}
which is however suppressed relative to the one in eq.~(\ref{eq:BoxNc}) by a
factor
$1/N_c^2$ and hence we neglect it. Therefore, we find that large--$N_c$
selects as the only operator contributing 
at the ``two--renormalon--chain'' level
the one in eq.~(\ref{eq:BoxNc}).
If we now Fierz the operator $\cT^{\mbox{\rm \small Box}}_2$, 
and use the identity 
\bea
\label{eq:Fierz} & & \mbox{\rm a}\ (\bar q \gamma^\m \frac{\la^a}{2} q)^2
 - \mbox{\rm b}\
(\bar q
\gamma^\m \gamma_5 \frac{\la^a}{2} q)^2 \equiv \nn \\ & &\frac{\mbox{\rm
a}-\mbox{\rm b}}{2} \left[
\sum_{a,b} \mid \bar q^a_L \gamma^\m q^b_L\mid^2 + (\mbox{\rm L} \rightarrow
\mbox{\rm R})\right] - 
\ 2 \ (\mbox{\rm a}+\mbox{\rm b}) \sum_{a,b} (\bar q^a_L q^b_R) \
(\bar q^b_R q^a_L) \,,
\eea 
we find that the result is almost like the four--fermion operators which
appear in the ENJL model of large--$N_c$ QCD~\cite{BBdR,Bij}. 
There is however a
very important difference, namely that the dimension--six 
$\cT^{\mbox{\rm \small Box}}$ operator 
which appears here is suppressed by the momentum
scale
$k_{E}^2$ and not by a constant scale like $\Lambda_{\chi}^2$. We shall
come back to this important issue in the next Section. Here we shall
limit ourselves to a numerical observation concerning the relative size of the
scalar--like coupling $G_S\sim 2(\mbox{\rm a}+\mbox{\rm b})$ and vector--like 
coupling
$G_V\sim ({\mbox{\rm a}-\mbox{\rm b}})/2$ which
appear in eq.~(\ref{eq:Fierz}). Plugging in the values $\mbox{\rm a}=10$, and
$\mbox{\rm b}=6$ obtained in eq.~(\ref{eq:BoxNc}) results in a ratio
$G_S/G_V=16$. For comparison we recall that a cut--off one--gluon
exchange yields $\mbox{\rm b}=0, G_S/G_V=4$, and the abelian version of
refs.~\cite{VZ94a,VZ94b}, which has $\mbox{\rm a}=4/3, \mbox{\rm b}=-3$,
would yield
$G_S/G_V=-20/13$, a negative value !. The phenomenologically favoured value of
ref.~\cite{BBdR} is $G_S/G_V\simeq 1$.

Let us come back now to the contribution of ``two renormalon chains'' to the
Adler function. The relevant diagram is the one depicted in Fig.~5, where the
black square stands for the insertion of the operator (\ref{eq:BoxNc}). As
already noticed in \cite{VZ94a,VZ94b} 
a subtlety arises at this point, for if one
wishes to keep the relations (\ref{eq:pivector}) 
and  (\ref{eq:t}), in order to
avoid double counting one has to multiply the right hand side of
eq.~(\ref{eq:pivector}) by a factor $1/2$. 
After Fierz--ing  the result in (\ref{eq:BoxNc}), only the vector part
contributes of course, and one obtains (recall eq.~(\ref{eq:pi1}) and the
fact that we are working in the large--$N_c$ limit)
\be
\label{eq:eighteen}
\Pi(q^2)= \frac{N_c}{16\pi^2}\ \left(\frac{N_c}{2}
\ \frac{1}{2\pi}\right)^2 \,\left( -\frac{1}{9}\right)
\ Q^2\int_{Q^2}^\infty
\frac{dk^2_E}{k_E^4}\ \left[\alphak\right]^2\ \log^2(k^2_E/Q^2)
\ee for the external vector vacuum polarization function, and 
\be
\label{eq:Atwochains}
\cA(Q^2)\mid^{\mbox{\rm \small two\ chains}}_{\mbox{\rm \small UV}}= \
\frac{N_c}{16\pi^2}\
\left(\frac{N_c}{2}\,\frac{1}{2\pi b_0}\right)^2\,\frac{2}{9}\,
\int_{0}^\infty  dw \ e^{- \frac{w}{b_0\alphaq}}\ \frac{w}{(1+w)^3}
\ee for the Adler function, where $w=b_0\alphaq \log(k^2_E/Q^2)$. 

At large
orders one finds
\be
\label{eq:Atwochainsexpanded}
\cA(Q^2)\mid^{\mbox{\rm \small two\ chains}}_{\mbox{\rm \small UV}}\sim \
\frac{N_c}{16\pi^2}\
\left(\frac{N_c}{2}\,\frac{1}{2\pi b_0}\right)^2\,\frac{1}{9}\, 
\sum_{\mbox{\rm \small Large}\ n} \left(- b_0 \alphaq \right)^n \ (n+1)! \ .
\ee 
It turns out that, both expressions (\ref{eq:Aonechainexpanded}) and
(\ref{eq:Atwochainsexpanded}) are comparable in the large--$N_c$ limit and of
the same order as the one--loop ``parton" graph. However,
(\ref{eq:Atwochainsexpanded}) is leading at large orders of perturbation
theory, i.e. for large $n$. In this sense one can view the
``two renormalon chain'' contribution as a selection of a subset of the whole
set of diagrams that are leading at large--$N_c$. This simplification may be 
welcome in the sense that, so far, even the leading contribution
at large--$N_c$
has proven itself to be already too difficult to deal with in QCD. We shall
comment on this again in the next Section.

\subsubsection{\normalsize{\bf Two Chains with the 
Gegenbauer Expansion Technique.}}
\label{subsubsec:TCGE}

We want to present another technique for calculating the ``two renormalon
chain'' contribution to the Adler function which uses directly the Feynman
diagrams generated by the ``amputated'' effective action in (\ref{eq:appr}).
Further limitation to leading contributions in the large--$N_c$ limit, and the
fact that we are only interested in the leading UV renormalon contribution
restricts the class of  possible diagrams with two chain exchanges to the
one in Fig.~6, with a Feynman gauge--like coupling i.e., in this diagram
only  the  $g^{\mu\nu}$ term in (\ref{eq:GBP}) is operational.

\vspace{5mm}

\centerline{\epsfbox{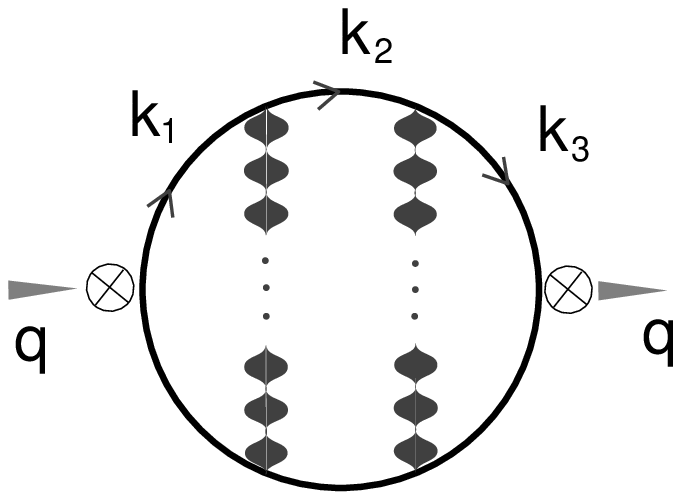}}
\vskip 1pc {{\bf Fig.~6} {\it Diagram representing 
the exchange of two effective
charges. A convenient routing of momenta to evaluate the contribution to the 
leading UV renormalon using the Gegenbauer expansion technique is 
indicated in the figure.}} 
\vspace*{8mm} 

\noi
With the routing
of momenta indicated in Fig.~6, the contribution to the vector two--point
function reads as follows:
\bea
\label{eq:gegens} 
\lefteqn{i\Pi^{\mu\nu}(q)  =  N_c\ \frac{N_c}{2}\
\frac{1}{2}\int\frac{d^4 k_1}{(2\pi)^4} 
\int\frac{d^4 k_2}{(2\pi)^4}\int\frac{d^4 k_3}{(2\pi)^4}\times }
\nn \\
 & &  
\ (4\pi)^2\ \int_{0}^{\infty} dz_1 e^{-z_1 \frac{1}{\alphamu}}
\int_{0}^{\infty} dz_2 e^{-z_2 \frac{1}{\alphamu}}
\frac{g_{\alpha\alpha'}(\mu^2)^{b_0 z_1}}{(k_2 -k_1 )^{2(1+b_0 z_1)}} 
\frac{g_{\beta\beta'}(\mu^2)^{b_0 z_2}}{(k_3 -k_2 )^{2(1+b_0 z_2)}}
\times \nn \\
 & & \frac{\tr
\left[\gamma^{\mu}\ks_{1} \gamma^{\alpha}\ks_{2}
\gamma^{\beta}\ks_{3}
\gamma^{\nu} (\ks_{3}
-\qs)\gamma^{\beta'}(\ks_{2}-\qs)\gamma^{\alpha'}(\ks_{1} -\qs)\right]}
{k_1^{2}k_2^{2}k_3^{2} (k_1 -q)^{2} (k_2 -q)^{2} (k_3 -q)^{2}}\,,
\eea 
where we have used the identity
\be
\alphak=\int_{0}^{\infty}dz
e^{-z\frac{1}{\alphamu}}\left(\frac{\mu^2}{k_{E}^2}\right)^{b_{0}z}\,.
\ee In order to calculate this integral, it is convenient to first 
introduce Dirichlet variables~\footnote{The change to the Dirichlet variables
$w$ and
$v$ automatically implements the convolution property of Laplace transforms.}:
\be
\label{eq:dirichlet}
w_{1}=b_{0}z_{1}\,,\qquad w_{2}=b_{0}z_{2}\,;\qquad
w=w_{1}+w_{2}\,,
\qquad v=\frac{w_{2}}{w}\,;
\ee 
and expand the denominators which depend on differences of momenta in
Gegenbauer polynomials~\footnote{Useful properties of Gegenbauer 
polynomials in
connection with Feynman diagram calculations can be found, e.g., in
ref.\cite{PT84} .}:
\be
\frac{1}{(k-p)^{2\lambda}}\ra\frac{-1}{\left( k_{E}^2 -2k_{E}p_{E}\cos\theta
+p_{E}\right)^ {2\lambda}}=
\nn
\ee
\be
\frac{-1}{k_{E}^2}\Theta\left(1-\frac{p_{E}}{k_{E}}\right)\sum_{n=0}^{\infty}
C_{n}^{\lambda}(\cos\theta)\left(\frac{p_{E}}{k_{E}}\right)^{n}+
\frac{-1}{p_{E}^2}\Theta\left(1-\frac{k_{E}}{p_{E}}\right)\sum_{n=0}^{\infty}
C_{n}^{\lambda}(\cos\theta)\left(\frac{k_{E}}{p_{E}}\right)^{n}\,,
\ee 
where the $C_{n}^{\lambda}(z)$, $[C_{n}^{1}(z)\equiv C_{n}(z)]$ are
Gegenbauer polynomials with the orthogonality property:
\be
\int d\Omega_{k}
C_{n}^{\lambda}(\hat{a}\cdot\hat{k})C_{m}^{\lambda}(\hat{b}\cdot\hat{k})=
\delta_{n,m}\frac{\lambda}{\lambda +n}C_{n}^{\lambda}(\hat{a}\cdot\hat{b})\,,
\ee with $C_{n}^{\lambda}(1)=\frac{\Gamma(n+2\lambda)}{n!\Gamma(2\lambda)}$;
and $ d\Omega_{k}$ the solid angular element resulting from
\be
\int d^{(4-\epsilon)}k\ra i\frac{2\pi^{(2-\epsilon/2)}}{\Gamma(2-\epsilon/2)}
\left(k_{E}\right)^{3-\epsilon}dk_{E}d\Omega_{k}\,.
\ee

The trace in the numerator of eq.~(\ref{eq:gegens}) also has to be expressed
in terms of powers of the Euclidean momenta $k_{1E}$, $k_{2E}$, $k_{3E}$,
$Q\equiv q_{E}$; and powers of $\cos\theta_{i0}\equiv\cos(\hat{k_{i}}\cdot 
\hat{q})$, $i=1,2,3$; and $\cos\theta_{ij}\equiv\cos(\hat{k_{i}}\cdot 
\hat{k_{j}})$. The Dirac trace results
in at most three powers of
$\cos\theta$'s; and therefore, from the orthogonality properties of the
Gegenbauer polynomial, only a few terms give non--zero contributions to the
solid angle integrals $\int d\Omega_{k_{1}}d\Omega_{k_{2}}d\Omega_{k_{3}}$.
With further restriction to those terms which can contribute to the leading UV
renormalon, one finds that only terms with two 
powers of $\qs$ and terms with no
powers of $\qs$ in the Dirac trace are relevant, with contributions
proportional to
$Q^{4}k_{2E}^2$ and $-\frac{2}{3}Q^{4}k_{2E}^2$ respectively. This results in
a contribution to $\Pi(q^2)$ from the leading UV renormalon generated by the
exchange of two chains:
\bea
\Pi(q^2) & = &
\frac{1}{3}\left(1-\frac{2}{3}\right)\frac{N_c}{16\pi^2}
\left(\frac{N_c}{2}\frac{1}{2\pi b_{0}}\right)^2
\int_{0}^{\infty}dw\,w\,e^{-w\frac{1}{b_{0}\alphamu}}\times \nn \\
 & & 
\int_{Q^2}^{\infty}\frac{dk_{2E}^2}{k_{2E}^2}
\left(\frac{Q^2}{k_{2E}^2}\right)^{1+w}
\int_{Q^2}^{k_{2E}^2}\frac{dk_{1E}^2}{k_{1E}^2}
\int_{Q^2}^{k_{2E}^2}\frac{dk_{3E}^2}{k_{3E}^2}
\left(\frac{Q^2}{\mu^2}\right)^{-w}\nn \\ & & \nn \\
 & = & \frac{N_c}{16\pi^2} \left(\frac{N_c}{2}\frac{1}{2\pi
b_{0}}\right)^2\,\frac{1}{9}\, 
\int_{0}^{\infty}dw\,w\,e^{-w\frac{1}{b_{0}\alphamu}}
\left(\frac{Q^2}{\mu^2}\right)^{-w}\int_{0}^{1}dx\, x^w \log^{2}\frac{1}{x}\,.
\eea    
The corresponding contribution to the Adler function, scaled to
$\mu^2=Q^2$, is then
\be
\cA (Q^2)_{\UV}^{\mbox{\rm \small two\ chains}}=\frac{N_c}{16\pi^2}
\left(\frac{N_c}{2}\frac{1}{2\pi b_{0}}\right)^2\,\left(-\frac{2}{9}\right)
\int_{0}^{\infty}dw\,e^{-w\frac{1}{b_{0}\alphaq}}\frac{1}{(1+w)^3}\,,
\ee in agreement with the OPE result in eq.~(\ref{eq:Atwochains}).

\vspace*{0.5cm} 

The final result of the leading UV renormalon contribution from the exchange of
one and two powers of the QCD effective charge to the Adler function
${\cal A}(Q^2)$; i.e., the sum of one and two renormalon chains, is then given
by the expression: 
\bea
\label{eq:VTPFUVRS} {\cal A}(Q^2) & = & 
\frac{N_c}{16\pi^2}\, \frac{N_c}{2}\,
\frac{1}{2\pi b_0}\,\frac{4}{9}\,\int_{0}^{\infty}dw
e^{-w\frac{1}{b_0\alpha (Q^2)}}
\nn \\
 & & \left\{ \frac{1}{(1+w)^2} -\frac{N_c}{2}\frac{1}{2\pi
b_0}\frac{1}{2}\frac{1}{(1+w)^3}\right\}\, ,
\eea  where the first term in the second line corresponds to the contribution
{}from the one renormalon chain, and the second term to the contribution from
the two renormalon chain.

\vspace*{1cm}
\subsection{The Pseudoscalar Two--Point Function}
\label{subsec:PTPF}

We shall next discuss the contribution of the leading UV renormalon to the
two--point function $\Psi_{5}(q^2)$ defined by eqs.~(\ref{eq:psifive}) and
(\ref{eq:axial}), calculated also at the level of one and two chains. Here
we shall first describe the calculations we have done using the Gegenbauer
expansion technique.  

\subsubsection{\normalsize {\bf One Chain with the Gegenbauer Expansion
Technique.}}
\label{subsubsec:OCGE}

There are three possible diagrams with a one renormalon chain contributing to
the two--point function $\Psi_{5}(q^2)$. Much the same as in the case of 
the vector two--point function, if we restrict ourselves to
the contribution from the leading UV renormalon, it is then sufficient to
consider the contribution from the one renormalon chain exchange in Fig.~7 with
a Feynman gauge--like coupling of the renormalon chain.

\vspace{5mm}

\centerline{\epsfbox{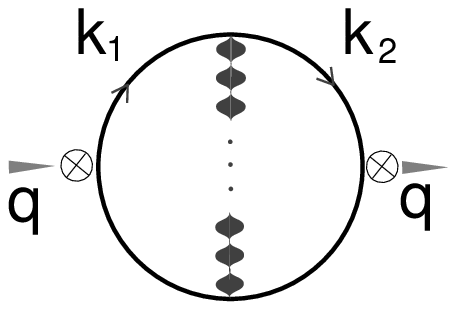}}
\vskip 1pc {{\bf Fig.~7} {\it Diagram representing 
the exchange of one effective
charge. A convenient routing of momenta to evaluate the contribution to the 
leading UV renormalon using the Gegenbauer expansion technique is the one
indicated in the figure.}} 
\vspace*{8mm} 

\noi With the routing of
momenta indicated in Fig~7, the contribution to
$\Psi_{5}(q^2)$ in large--$N_c$ is then as follows:
\bea
\label{eq:psi1}
\Psi_{5}(q^2) & = & -i(m_{u}+m_{d})^2 N_c\,\frac{N_c}{2}
\int\frac{d^{4}k_{1}}{(2\pi^4)}\int \frac{d^{4}k_{2}}{(2\pi^4)}
\frac{-i\left(-ig_{\eff}\left[-(k_{1}-k_{2})^2\right]\right)^2}{(k_{1}-k_{2})^2
+i\epsilon}\times
\nn \\
 & & 
\frac{\tr\left[ i\gamma_{5}\ks_{1}\gamma^{\mu}\ks_{2}i\gamma_{5}(\ks_{2}-\qs)
\gamma_{\mu}(\ks_{1}-\qs)\right]}{\left[k_{1}^2 +i\epsilon\right]
\left[k_{2}^2 +i\epsilon\right]\left[\left(k_{1}-q\right)^2 +i\epsilon\right]
\left[\left(k_{1}-q\right)^2 +i\epsilon\right]}
\,.
\eea 
Using the identity
\be
\frac{-i\left(-ig_{\eff}\left[-(k_{1}-k_{2})^2\right]\right)^2}{(k_{1}-k_{2})^2
+i\epsilon}=
-i\frac{4\pi}{\mu^2}\int_{0}^{\infty}dz\,e^{-z\frac{1}{\alphamu}}
\left(\frac{\mu^2}{(k_{1E}-k_{2E})^2}\right)^{1+b_{0}z}\,;
\ee 
and expanding the denominators which depend on differences of momenta in
Gegenbauer polynomials as in subsection~\ref{subsubsec:OCGE}, one obtains the
expression
\bea
\label{eq:psi2}
\lefteqn{\Psi_{5}(q^2)  =  \frac{(m_{u}+m_{d})^2}{\mu^2}
\frac{N_c}{16\pi^2}\,\frac{N_c}{2}\frac{1}{2\pi b_0}\int_{0}^{\infty}dw\,
e^{-w\frac{1}{b_{0}\alphamu}}\left(\frac{\mu^2}{Q^2}\right)^{1+w}\times}\nn\\
 & &
\int_{Q^2}^{\infty}\frac{dk_{1E}^2}{k_{1E}^2}
\left(\frac{Q^2}{k_{1E}^2}\right)^{1+w}\int_{Q^2}^{k_{1E}^2}
\frac{dk_{2E}^2}{k_{2E}^2}\int d\Omega_{k_{1}}\int d\Omega_{k_{2}}
\sum_{l}C_{n_{l}}^{(1+w)}(\cos\theta_{12})
\left(\frac{k_{2E}}{k_{1E}}\right)^{l}\times \nn
\\
 &  & 
\sum_{n_{1}}C_{n_{1}}(\cos\theta_{10})\left(\frac{Q}{k_{1E}}\right)^{n_{1}}
\sum_{n_{2}}C_{n_{2}}(\cos\theta_{20})\left(\frac{Q}{k_{2E}}\right)^{n_{2}}
\times \nn\\
 & & 16
\left[ k_{1E}^2 k_{2E}^2 -  k_{1E}^2 k_{2E}Q\cos\theta_{20} - k_{2E}^2
k_{1E}Q\cos\theta_{10} +
k_{1E}k_{2E}Q^{2}\cos\theta_{10}\cos\theta_{20}\right]\,,
\eea where $w=b_{0}z$ and
$\cos\theta_{i0}\equiv\cos(\hat{k_{i}}\cdot\hat{q})$ for $i=1,2$ and
$\cos\theta_{12}\equiv\cos(\hat{k_{1}}\cdot\hat{k_{2}})$. 

The leading UV renormalon contribution in eq.~(\ref{eq:psi2}) comes solely 
from the $l=0$ term and the
$k_{1E}k_{2E}Q^{2}\cos\theta_{10}\cos\theta_{20}$ term resulting from the Dirac
trace. The angular integrals can then be trivially done, with the result
\bea
\label{eq:psi3}
\Psi_{5}(q^2)\vert_{\UV}^{\mbox{\rm \small one\ chain}} & = &
\frac{(m_{u}+m_{d})^2}{\mu^2}
\frac{N_c}{16\pi^2}\,\frac{N_c}{2}\frac{1}{2\pi b_0}\int_{0}^{\infty}dw\,
e^{-w\frac{1}{b_{0}\alphamu}}\left(\frac{\mu^2}{Q^2}\right)^{1+w}\times\nn\\
 & &  
\int_{Q^2}^{\infty}\frac{dk_{1E}^2}{k_{1E}^2}
\left(\frac{Q^2}{k_{1E}^2}\right)^{1+w}\,4 Q^4\,\log\frac{k_{1E}^2}{Q^2}\,;
\eea and therefore
\be
\Psi_{5}(q^2)\vert_{\UV}^{\mbox{\rm \small one\ chain}}
 = (m_{u}+m_{d})^2\mu^2\frac{N_c}{16\pi^2}\frac{N_c}{2}
\frac{1}{2\pi b_0}\int_{0}^{\infty}dw\,
e^{-w\frac{1}{b_{0}\alphamu}}\left(\frac{Q^2}{\mu^2}\right)^{1-w}
\frac{4}{(1+w)^2}\,.
\ee  
Taking two derivatives with respect to $Q^2$ and scaling $\mu^2$ at
$\mu^2=Q^2$ we get the corresponding contribution to the
pseudoscalar--pseudoscalar $\cP (q^2)$ correlation function:
\be
\label{eq:pfunct1}
\cP(q^2)\vert_{\UV}^{\mbox{\rm one\ chain}}= \frac{N_c}{16\pi^2}\frac{N_c}{2}
\frac{1}{2\pi b_0}\int_{0}^{\infty}dw\, e^{-w\frac{1}{b_{0}\alphaq}}
\frac{8}{(1+w)^2}\,.
\ee

\subsubsection{\normalsize {\bf Two Chains with the Gegenbauer Expansion
Technique.}}
\label{subsubsec:TCGEP}

As in the case of the two--chain evaluation of the Adler function in
subsection~{\ref{subsubsec:TCGE}, the relevant 
diagram is the one in Fig.~6 with
now external pseudoscalar sources, and with a 
Feynman gauge--like coupling in the
renormalon chains. With the routing of momenta indicated in Fig.~6 the
contribution to
$\Psi_{5}(q^2)$ can be readily written as follows
\bea
\lefteqn{\Psi_{5}(q^2)= i(m_u +m_d)^2\,N_c\,\left(\frac{N_c}{2}\right)^2\,
i\frac{\pi^2}{(2\pi)^4}\,i\frac{\pi^2}{(2\pi)^4}\,i\frac{\pi^2}{(2\pi)^4}\,
\int_{Q^2}^{\infty}\frac{dk_{2E}^2}{k_{2E}^2}
\int_{Q^2}^{k_{2E}^2}\frac{dk_{1E}^2}{k_{1E}^2}
\int_{Q^2}^{k_{2E}^2}\frac{dk_{3E}^2}{k_{3E}^2}}\nn \\ 
 & & 
\times\int d\Omega_{k_1}\int d\Omega_{k_2}\int d\Omega_{k_3}\,
\tr\left[i\gamma_{5}\ks_{1}\gamma^{\mu}\ks_{2}\gamma^{\nu}\ks_{3}i\gamma_{5}
(\ks_{3}-\qs)\gamma_{\nu}(\ks_{2}-\qs)
\gamma_{\mu}(\ks_{1}-\qs)\right]\,\nn \\
 & & 
\times\left(\frac{4\pi}{b_0}\right)^2\int_{0}^{\infty}dw_{1}
e^{-w_{1}\frac{1}{b_{0}
\alphamu}}\int_{0}^{\infty}dw_{2} 
e^{-w_{2}\frac{1}{b_{0}\alphamu}} \nn\\
 & & 
\times\Sigma_{n_{1}}C_{n_{1}}(\cos\theta_{10})
\left(\frac{Q}{k_{1E}}\right)^{n_{1}}
\Sigma_{n_{2}}C_{n_{2}}(\cos\theta_{20})
\left(\frac{Q}{k_{2E}}\right)^{n_{2}}
\Sigma_{n_{3}}C_{n_{3}}(\cos\theta_{30})
\left(\frac{Q}{k_{3E}}\right)^{n_{3}} \nn \\
& & 
\times\frac{(\mu^2)^{w_{1}+w_{2}}}{(k_{2E}^2)^{1+w_{1}+w_{2}}}\Sigma_{l_{1}}
C_{l_{1}}^{1+w_{1}}
(\cos_{12})\left(\frac{k_{1E}}{k_{2E}}\right)^{l_{1}}
\Sigma_{l_{2}}C_{l_{2}}^{1+w_{2}}
(\cos_{32})\left(\frac{k_{3E}}{k_{2E}}\right)^{l_{2}}\,.
\eea The configurations with $k_{2E}\le k_{1E}$ in the Gegenbauer expansion
have been dropped because they do not contribute to the leading UV renormalon. 
The evaluation of the trace gives the rather simple result
\be
\tr\left[...\right] =-64\,
\left(k_{1E}^2-k_{1E}Q\cos\theta_{10}\right)
\left(k_{2E}^2-k_{2E}Q\cos\theta_{20}\right)
\left(k_{3E}^2-k_{3E}Q\cos\theta_{30}\right)\,.
\ee

To proceed further we introduce Dirichlet variables, as in
eqs.~(\ref{eq:dirichlet}). The integral over the variable $v$ can be trivially
done; and the contribution to the leading UV renormalon comes only from the
terms with $l_{1}=l_{2}=n_{2}=0$, and $n_{1}=n_{3}=1$. The integral then
simplifies a lot, with the result
\bea
\frac{1}{(m_u+m_d)^2}\Psi_{5}(q^2) & = & \frac{N_c}{16\pi^2}
\,\left(\frac{N_c}{2}\right)^2\,\frac{4 Q^2}{(2\pi b_{0})^2}\,
\int_{0}^{\infty}dw\,w\, e^{-w\frac{1}{b_{0}\alphamu}}\times \nn \\ & &
\int_{Q^2}^{\infty}\frac{dk_{2E}^2}{k_{2E}^2}\left(\frac{Q^2}{k_{2E}^2}\right)
^{1+w}\,\log^{2}\frac{k_{2E}^2}{Q^2}\left(\frac{\mu^2}{Q^2}\right)^w\, \nn \\ &
= & \frac{N_c}{16\pi^2} \left(\frac{N_c}{2}\right)^2
\frac{4\mu^2}{(2\pi b_{0})^2}
\int_{0}^{\infty}dw e^{-w\frac{1}{b_{0}\alphamu}}\frac{2w}{(1+w)^3}
\left(\frac{Q^2}{\mu^2}\right)^{1-w}. 
\eea

Taking two derivatives with respect to $Q^2$ and scaling $\mu^2$ at
$\mu^2=Q^2$ we get as a final result for the two--chain contribution to the
pseudoscalar--pseudoscalar $\cP (q^2)$ correlation function the expression:

\be
\label{eq:2chaingege}
\cP(q^2)\vert_{\UV}^{\mbox{\rm \small two\ chains}}=
\frac{N_c}{16\pi^2}\,\left(\frac{N_c}{2}
\frac{1}{2\pi b_{0}}\right)^2\,\left(-16\right)
\int_{0}^{\infty}dw\,e^{-w\frac{1}{b_{0}\alphaq}}\frac{1}{(1+w)^3}\,.
\ee

\subsubsection{\normalsize  {\bf The One and Two Chain Calculations with the
OPE Technique.}}
\label{subsubsec:OTCOPE}

The calculation of the pseudoscalar two--point function with the OPE technique
is done along the same lines as the one described earlier for the vector
two--point  function. Recalling the definitions (\ref{eq:Dirac}),
(\ref{eq:axial}),  (\ref{eq:psifive}), (\ref{eq:psifiveder}) one finds that 
\be
\label{eq:psi5ope}
\Psi_5(q^2)= -\ \frac{1}{2} \int_{k^2_E\ge Q^2} \frac{d^4k}{(2\pi)^4}\
\frac{4\pi \alphak}{k^2}\ \langle p\mid \frac{\cT}{k^2} \mid p\rangle
\ee where $\cT$ is the same as in eq.~(\ref{eq:t}), and $\mid p\rangle$ is the
pseudoscalar ``particle" annihilated by the external field $p(x)$ of
eq.~(\ref{eq:Dirac}). Contrary to what happened for the vector two--point
function, now at the level of the ``one renormalon chain'' one does find {\it a
priori} a dimension--four operator in 
the expansion of the operator $\cT$ which,
in the  large--$N_c$ limit, reads
\be
\cT_{dim. 4} = 3 N_c \ \bar q(0)i\gamma_5 q(0)\ p(0)\,;
\ee as well as some new dimension--six operators
\be
\label{eq:taxial}
\cT_{dim. 6}= \frac{N_c}{k^2}\ \bar q(0)i\gamma_5 q(0)\ 
\left[ 2\ \partial^2 p(0) + 4\ \ p(0)^3\right]
\ee besides, of course, the 
term $D^\al G_{\al \m}$ of eq.~(\ref{eq:tone}) which,
being subleading in $N_c$, we shall not consider any further.
Dimension four operators do not contribute to $\cP(Q^2)$~\cite{VZ94a,VZ94b}. 
As for the contribution of the
dimension six operators of eq.~(\ref{eq:taxial}), the relevant diagram is the
one analogous to the vertex form factor depicted in Fig.~4 and yields
\be
\label{eq:Ponechaink}
\cP(Q^2)\vert_{\UV}^{\mbox{\rm \small one\ chain}}
 = \frac{4N_c^2}{(4\pi)^3}\ Q^2 \int_{Q^2}^\infty
\frac{dk_E^2}{k^4_E}\ \alphak \ \log(k_E^2/Q^2)
\ee which, with the change of variable $w=b_0\alphaq \log(k_E^2/Q^2)$, reads
\be
\label{eq:Ponechainw}
\cP(q^2)\vert_{\UV}^{\mbox{\rm \small one\ chain}} = 
\frac{N_c}{16\pi^2}\frac{N_c}{2}
\frac{1}{2\pi b_0}\int_{0}^{\infty}dw\, e^{-w\frac{1}{b_{0}\alphamu}}
\frac{8}{(1+w)^2}\,,
\ee in agreement with the result (\ref{eq:pfunct1}) obtained with the 
Gegenbauer expansion technique. 

When considering two renormalon chains, the four--fermion operators of 
eq.~(\ref{eq:BoxNc}) also contribute to $\cP(Q^2)$ through  diagrams 
analogous to the ones in Fig.~5, with the result
\be
\label{eq:Ptwochainsk}
\cP(Q^2)= \frac{8N_c^3}{(4\pi)^4}\ Q^2 \int_{Q^2}^\infty
\frac{dk_E^2}{k^4_E}\ \left[\alphak\right]^2 \ \log^2(k_E^2/Q^2)\,,
\ee or with the above change of variables 
\be
\label{eq:Ptwochainsw}
\cP(Q^2)_{\UV}^{\mbox{\rm \small two\ chains}} =
 \ \frac{N_c}{16\pi^2}
\left(\frac{N_c}{2}\frac{1}{2\pi b_0}\right)^2\,16 \int_0^\infty dw\
e^{-\frac{w}{b_0\alphaq}}
\ \frac{w}{(1+w)^3}\ ,
\ee also in agreement with the result obtained in eq.~(\ref{eq:2chaingege})

\vspace*{0.5cm} 

The final result of the leading UV renormalon contribution from the exchange
of one and two powers of  the QCD effective charge to the function
${\cal P}(Q^2)$; i.e., the sum of one and two renormalon chains, is then given
by the expression: 
\bea
\label{eq:PTPFUVRS} {\cal P}(Q^2) & = & 
\frac{N_c}{16\pi^2}\, \frac{N_c}{2}\,
\frac{1}{2\pi b_0}\,8\int_{0}^{\infty}dw e^{-w\frac{1}{b_0\alpha (Q^2)}}
\nn \\
 & & \left\{ \frac{1}{(1+w)^2} -\frac{N_c}{2}\frac{1}{2\pi
b_0}\,2\,\frac{1}{(1+w)^3}\right\}\,.
\eea  where the first term in the second line corresponds to the contribution
{}from the one renormalon chain, and the second term to the contribution from
the two renormalon chains.

\subsubsection{\normalsize {\bf Comments on the two--point function
calculations.}}
\label{subsubsec:Comments} 

We shall conclude this Section with various comments based on the previous
calculations.

\begin{itemize}
\item {\it Large--$N_c$ equivalences.}

 Recall that $b_{0}\equiv
\frac{-\beta_{1}}{2\pi}$, and
$\beta_{1}=
\frac{-11}{6}N_c +\frac{1}{3}n_f$. Since we are working to leading order in
the $1/N_c$--expansion, we have the following equivalences:
\be
\frac{1}{2\pi b_{0}}C_{F}\rightarrow \frac{1}{2\pi b_{0}}\frac{N_c}{2}\ra
\frac{3}{11}\,.
\ee
We  keep the second algebraic form explicitly in most of the
results because it shows better the origin of the various factors and the
fact that we are neglecting non--leading contributions.

\item {\it Large order behaviour.}

The large order behaviour in perturbation
theory of the Adler function and the pseudoscalar correlation function ${\cal
P}(Q^2)$  can be read off from the results in eqs.~(\ref{eq:VTPFUVRS}) and
(\ref{eq:PTPFUVRS}) above. They are:
\be 
\label{eq:asymptone}
{\cal A}(Q^2)\vert_{\mbox{\rm UV}}\sim \frac{N_c}{16\pi^2}
\frac{4}{33} (-1)^{n-1}n!\left(1-\frac{3}{44}n\right)
(b_{0}\alpha_{s}(Q^2))^{n}\,,
\ee
\noi  and
\be 
\label{eq:asymptotwo}
{\cal P}(Q^2)\vert_{\mbox{\rm UV}}\sim 
\frac{N_c}{16\pi^2}\frac{24}{11}(-1)^{n-1}n!\left(1-\frac{6}{22}n\right)
(b_{0}\alpha_{s}(Q^2))^{n}\,.
\ee
As already discussed above, these results show that,
asymptotically, the effect of the two UV renormalon chains is leading with
respect to the one UV renormalon chain. Notice also that the overall
coefficient of the pseudoscalar ${\cal P}(Q^2)\vert_{\mbox{\rm
UV}}$--function is an order of magnitude larger than the corresponding one in
the Adler
${\cal A}(Q^2)\vert_{\mbox{\rm UV}}$ --function. Taking the 
results (\ref{eq:asymptone},\ref{eq:asymptotwo}) at face value, 
one would infer that the dominance of the two
renormalon chain in the pseudoscalar correlation function happens at earlier
$n$ values  than in the vector correlation function.

\item {\it The expansion in terms of local operators.}

As we have seen from the previous calculations, one and two chains of bubbles
generate at a large Euclidean scale
$k_{E}$ a set of local operators (on the scale $1/k_{E}$) of increasing
dimensionality, i.e., 
\be
\label{eq:star}
\cT = \sum_{i\ge 6} \frac{c_i(k_{E})}{(k_{E})^{n_{i}-4}}\ \cO_i (0)\ ,
\ee
where $n_i=$ dimension of the operator $\cO_i (0)$. In this equation 
$c_i(k_{E})$ are the corresponding 
Wilson coefficients to be obtained as a power
series in $\alphak$. The crucial observation, already made in
refs.~\cite{VZ94a,VZ94b}, is that the physics starts with operators $\cO_i$ of
dimension not smaller than six. As explained in \cite{VZ94b} operators of
dimension smaller than six have their contribution to the two--point functions
buried in the infinite renormalization constant already present at the parton
level and that drops out once enough derivatives with respect to the external
momenta are taken in the corresponding two--point function.

Furthermore the expansion of eq.~(\ref{eq:star}) is meaningful, in the sense
that contributions from operators
$\cO_i$ of higher dimension give rise to  singularities in the Borel plane
further away from the origin, i.e. their $n-$th coefficient in the asymptotic
perturbative expansion is more and more suppressed.

\item {\it Diagrams with increasing complexity.}

As one considers contributions
coming from diagrams with an increasing complexity, i.e. more and more chains,
the different Wilson coefficients
$c_i(k_{E})$ in (\ref{eq:star}) have more and more powers of
$\alphak$. One can easily see that integrals like 
\be
\label{eq:manychains}
I_{p}=Q^2\ \int_{Q^2}^\infty \frac{dk^2}{k^4}\ 
[\alphak]^p \ \log^2(k^2/Q^2) \ ,
\ee
which will appear in the two--point functions (\ref{eq:eighteen}) and 
(\ref{eq:Ptwochainsk}), give a contribution to the leading term in the
perturbative asymptotic expansion that is independent of $p$ and,
consequently, powers of $\alphak$ are not suppressed. To see this just use

\bea
\label{eq:expansion}
[\alphak]^p =& & \frac{[\alphaq]^p}{\left( 1 + b_0 \alphaq \log(k^2/Q^2)
\right)^p} = \nn \\
& &[\alphaq]^p \sum_{n=0}^\infty \frac{\left( - b_0\alphaq
\log(k^2/Q^2)\right)^n }{n!}\ \frac{\Gamma(n+p)}{\Gamma(p)}
\eea
and insert it in eq.~(\ref{eq:manychains}). After the $k$ integration one
finds
\be
I=\frac{1}{\Gamma(p)} \sum_{n=0}^\infty \frac{(- b_0)^n}{n!} \alphaq^{n+p}
\ \Gamma(n+p) \Gamma(n+3)\ .
\ee

At large $n$ one can shift $n \rightarrow n-p$ to obtain
\be
I_{p}=\frac{1}{\Gamma(p) (-b_0)^p} \sum_{\mbox{\rm Large\ n}} 
(- b_0 \alphaq)^n (n+1)! \left[\frac{\Gamma(n+3-p)}
{n (n+1) \Gamma(n+1-p)}\right] \ .
\ee
One can see that the expression between square  
brackets goes to unity at large $n$,
independent of $p$. The factor $1/\Gamma(p)$ out front affects the residue
but not the position of the singularity on the Borel plane at $-1/b_0$. In
fact one does not expect any suppression from large values of $p$ in the
final contribution, as the coefficient accompanying the $\alphak^p$ term a
priori may behave as  $p!$ \cite{VZ94a,VZ94b}. This effectively 
renders the residue of the
UV renormalon incalculable.

Our discussion has been simplified  by not including the effect of anomalous
dimensions of the operators $\cO_i$ of eq.~(\ref{eq:star}). They introduce
contributions such as $[\alphaq \log(k^2/Q^2)]^m$ for an arbitrary power $m$
in the integral (\ref{eq:manychains}). Taking into account anomalous
dimensions makes the analysis more cumbersome but does not change the main
conclusion about the dominance of four--fermion operators 
(see refs. \cite{VZ94b,BS96} for details).

\item{\it What message does one learn from these calculations?}

For the class of QCD contributions which are obtained with the ``amputated''
generating functional of eq.~(\ref{eq:appr}) we argue that, when the
momentum transfer in the {\it effective charges} exchanged are very large, only
two types of effective local operators can appear, 
depending on whether the large
momentum flows to the external vertex or not. The local operators are
respectively vertex--like operators which connect two fermions to the
relevant external source (like the case illustrated in Fig.~4,) and
four--fermion box--like operators (like the case illustrated in Fig.~5.) 
As we have seen the contribution from the dimension $d=6$  four--fermion
operators dominates because of the two powers 
of $\log\frac{k_E^2}{Q^2}$ factors
{}from each fermion loop. In the large
$N_c$ limit, the only $d=6$ four--fermion operators which are
allowed by the chiral symmetry properties of perturbation theory are the two
operators of the ENJL model. Four--fermion tensor  operators like
\be
\sum_{i,j={\mbox{\rm \small flavour}}} \left(\bar \psi_L^i \sigma^{\mu\nu}
\psi_R^j\right)\ \left(\bar \psi_R^j \sigma_{\mu\nu} \psi_L^i\right)\ ,
\ee 
which is chirally invariant and where colour is summed over within
the parentheses, can be seen to vanish identically for instance by employing 
the identity $\sigma^{\mu\nu} (1\pm \gamma_5) = 
\sigma^{\mu\nu} \pm \frac{i}{2} \epsilon^{\mu\nu\rho\lambda} \sigma_{\rho
\lambda}$ \ \footnote{We thank J. Bijnens for pointing this out to us.}.

In theories like QED, where the UV renormalon can be interpreted as the signal
that non--perturbative contributions must exist, there are indications from
lattice simulations that four--fermion operators may indeed play an important
r\^{o}le in non--perturbative dynamics
\cite{Az}. In QCD it is difficult to imagine how any non--perturbative
dynamics may emerge from the study of UV renormalons 
since they involve only very
high momenta. Our point of view is that the richness revealed by the UV
renormalon structure may perhaps be used as well in QCD to learn about generic
features of non--perturbative physics provided one studies a regime where there
is some {\it infrared momenta} involved. The only possible choice in the
two--point functions which we are studying seems 
to be the regime where the {\it external momenta}
$Q^2$ is taken to be very small instead of very large as is usually assumed.
One can think of many physical processes which involve integrals over the
external field Euclidean momenta $Q^2$ of these two--point functions all the
way down to zero. The hadronic vacuum polarization contribution to the
anomalous magnetic moment of the muon is a	 well known example. (See e.g.
ref.\cite{deR94} and references therein.) It is this ``atypical'' situation of
{\it small external momenta} which we next want to explore.

\end{itemize}

\newpage
\setcounter{equation}{0}

\section{The fate of UV--Renormalons in Two--Point Functions at low $Q^2$
values.}
\label{sec:UVLow}

We wish now to discuss what happens to the UV renormalon
contributions to two--point functions when the external Euclidean momenta
$Q^2$ becomes smaller and smaller. We shall take as a starting point the
contribution of two chains to the vector two--point function in
eq.~(\ref{eq:eighteen}). One can readily obtain  
{}from it the following leading contribution
to the Adler function  
\be
\Pi(q^2)= \frac{N_c}{16\pi^2}\ \left(\frac{N_c}{2}
\ \frac{1}{2\pi}\right)^2 \,\left( -\frac{1}{9}\right)
\  Q^2\int_{Q^2}^\infty
\frac{dk^2_E}{k_E^4}\ \left[\alphak\right]^2\ \log^2(k^2_E/Q^2)\ .
\ee
This integral is well defined as long as $Q^2> \Lambda_L^2$; however, for
$Q^2< \Lambda_L^2$ there is a pole in the integration range. Let us examine
this more closely. 
With the change of variables $w=\alpha(\m) b_0 \log(k_E^2/Q^2)$,
we can rewrite eq.~(\ref{eq:Atwochains}) as follows
\be
\label{eq:a}
\cA(Q^2)= - \frac{N_c}{16\pi^2}\
\left(\frac{N_c}{2}\,\frac{1}{2\pi b_0}\right)^2 \left( \frac{1}{9}\right)
\frac{1}{b_0\alpham}
\int_0^\infty  dw \ e^{- \frac{w}{b_0\alpham}}\ \frac{w^2}{(1+w_Q+w)^2}
\ee
where $\alpham$ is the running coupling constant defined at a scale $\mu^2>
 \Lambda_L^2$, where a perturbative expansion in powers of  $\alpham$
makes sense a priori. In this equation 
\be 
w_Q\equiv \alpha(\m) b_0
\log(Q^2/\mu^2)\,,
\ee
and in the range $Q^2\le \Lambda_L^2$ one finds that
$w_Q\le -1$ so that, indeed, the above integrand has a double pole at
$w=-1- w_Q$, i.e. within the integration range.\footnote{One can easily check
that if $Q^2\ge \Lambda_L^2$ one may choose $\m =Q^2$ and integrate by parts
to recover eq.~(\ref{eq:Atwochains}).} Consequently this integral will be
ambiguous depending on how one goes around 
the double pole. This ambiguity turns
out to be $\mu$--independent and reads
\be
\label{eq:b}
\delta \cA(Q^2)_{Q^2<\Lambda_{L}^2} = -\ \cK \frac{N_c}{16\pi^2}\
\left(\frac{N_c}{2}\,\frac{1}{2\pi b_0}\right)^2 \ 
\frac{Q^2}{\Lambda_L^2}\ 
\left(\frac{1}{9} 
\log^2\frac{\Lambda_L^2}{Q^2} - 2 \log \frac{\Lambda_L^2}{Q^2} \right)
\ee
where $\cK$ is an unknown constant parameterizing the ambiguity and of
$\cO (N_c^0)$. 

We think this is quite a remarkable result! The ambiguity turns out to be of
the same type as the insertion in the vector current two--point function
$\Pi(Q^2)$ of the {\it local} $d=6$ composite operator
\be
\label{eq:c}
\cO = - \frac{8\pi^2 G_V}{N_c \Lambda_{\chi}^2} 
\sum_{\mbox{\rm \small a,b=flavour}} 
\left[ (\bar q^a_L \gamma^\m q^b_L)\ (\bar q^b_L \gamma_\m q^a_L) 
+ (\mbox{\rm L} \rightarrow \mbox{\rm R}) \right]
\ee
provided one interprets $\Lambda_L^2$ as the momentum cutoff in the loops
and $\Lambda_{\chi}\approx \Lambda_L$.
This is one of the four--fermion operators appearing in the ENJL model
\cite{BBdR,Bij} with the identification 
\be
4 G_V \equiv - \cK\ \left(\frac{1}{2\pi b_{0}}\frac{N_c}{2}\right)^2\ .
\ee
The floating scale $k_{E}^2$ in eq.~(\ref{eq:BoxNc}) 
has now turned into a hard
scale
$\Lambda_L^2$.

We can repeat the same calculation starting from the integral in
eq.~(\ref{eq:ten}) for the contribution of the one--chain renormalon. The
ambiguity is again $\m$--independent but now it has one power less of the  
$\log \frac{\Lambda_L^2}{Q^2}$:
\be
\label{eq:d}
\delta \Pi(Q^2)\mid_{Q^2<\Lambda_{L}^2}^{\mbox{\rm \small one\ chain}} 
\sim \frac{N_c^2}{b_0}
\frac{Q^2}{\Lambda_L^2} \log \frac{\Lambda_L^2}{Q^2}
\ee
and consequently this ambiguity is subleading with respect to that of
eq.~(\ref{eq:b}) in the low--$Q^2$ limit. In Section 
\ref{sec:UVTPF} we have seen that, for
$Q^2$ large, the two--chain renormalon dominates  over the one--chain one at
large orders of perturbation theory . What we see now is that, for
$Q^2$ small,  a similar
dominance effect appears as well, and it manifests itself  in the form of an
extra
$\log
\frac{\Lambda_L^2}{Q^2}$ in the order of the ambiguity . The ambiguity in
(\ref{eq:d}) can also be reproduced by the insertion of a {\it local}
$d=6$ operator, which in this case is 
\be
\label{eq:e}
\sim \frac{N_c}{b_0 \Lambda_L^2} \ \bar q(x)\gamma^\m
\partial^{\al} F_{\al \m}(x) q(x)
\ee
with $F_{\al \m}(x)$ being the field strength tensor of the external vector
field $v_\mu (x)$. The normalization out front is of $\cO(N_c^0)$ but is
ambiguous. The
operator in eq.~(\ref{eq:e}) again coincides with the effective operator at
the scale $k_{E}$ of eq.~(\ref{eq:tone}) but with the important difference that
now there is a constant $\Lambda_L^2$ scale in the denominator.

If one {\it defines} 
\be
\label{eq:f}
\alphaq\equiv \frac{\alpham}{ \Big(1 + b_0 \alpham \log(Q^2/\mu^2)\Big)}
\ee
for $Q^2<\Lambda_L^2$, i.e. if one uses analytic continuation, expressions such
as eq.~(\ref{eq:a}) acquire a more familiar form. Notice that the $\alphaq$
defined in this way becomes negative for 
$Q^2<\Lambda_L^2$. By making the change
of variables 
\be
w'= - \ w \ \frac{\alphaq}{\alpham}
\ee
one can check that eq.~(\ref{eq:a}) becomes 
\be
\label{eq:g}
\cA(Q^2)= - \ \frac{N_c^3}{9 (4\pi)^4 b_0^2}\ \ \frac{1}{b_0\mid \alphaq\mid }
\int_0^\infty  dw' \ e^{- \frac{w'}{b_0 \mid \alphaq\mid }}\ 
\frac{w'^2}{(1 - w')^2}\ ,
\ee
and the relevant expansion parameter is now $\mid \alphaq\mid$, the absolute
value of $\alphaq$.  In the ordinary situation where $Q^2> \Lambda_L^2$ one
can follow the same steps and show that a plus sign appears instead in the
denominator, i.e. it is of the form $(1+w')^2$, as it should be. Of course in
this case eq.~(\ref{eq:f}) is not merely a definition but the true solution of
the renormalization group equation. 

\vspace*{0.5cm}

We shall next argue that it is not so surprising to find that UV renormalons
become real singularities when $Q^2\le
\Lambda_{L}^2$. It is helpful for this purpose to look at the
$Q^2$--$k_{E}^2$ plot in Fig.~8.

\vspace{5mm}

\centerline{\epsfbox{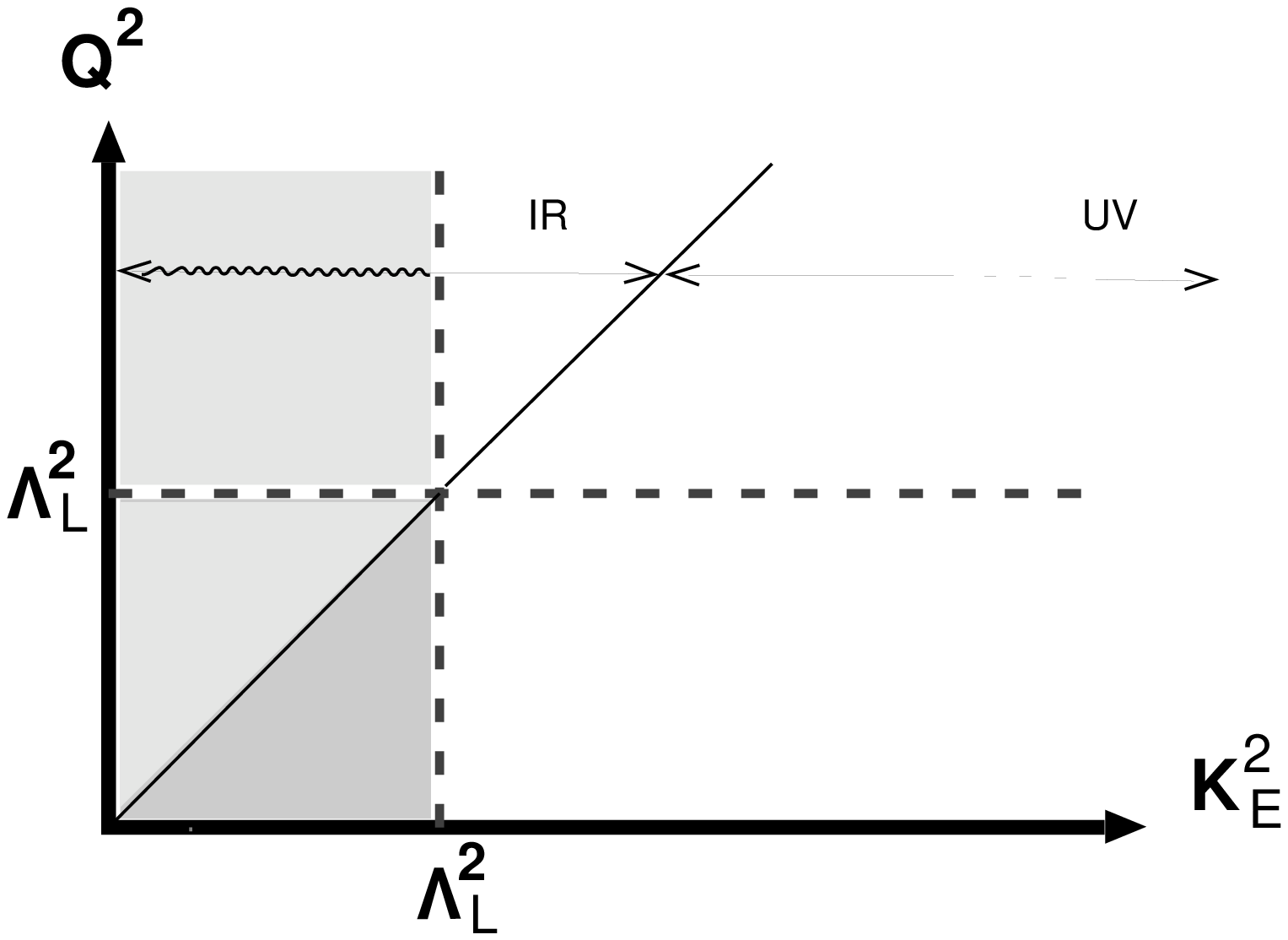}}
\vskip 1pc {{\bf Fig.~8} {\it Integration regions of the virtual Euclidean
momentum $K_{E}^2$ for a fixed Euclidean external momenta $Q^2$.}} 
\vspace*{8mm} 

\noi
The integration regions $0\le k_{E}^{2}\le Q^2$ and 
$Q^2 \le k_{E}^2\le\infty$ shown in Fig.~8 are at the origin of the appearance
of IR renormalons and UV renormalons in two--point functions, respectively. 
(Fig.~8 corresponds to the precise case of the one--chain
renormalon, but it can be easily generalized to an arbitrary number of chains
by introducing one more dimension for each renormalon chain.) For reference,
we also show in this plot the lines
$Q^2=\Lambda_{L}^2$ and
$k_{E}^2=\Lambda_{L}^2$. In the conventional study of IR renormalons, the
external Euclidean momenta is always chosen to be
$Q^2\gg\Lambda_{L}^2$;
however, regardless of how large
$Q^{2}$ is taken, there will always be a region in the virtual
$k_{E}^2$ integration which is below $\Lambda_{L}^2$ and which leads to
IR renormalon poles in the Borel plane. As discussed in Section 
\ref{sec:Eff} their
appearance is welcome because they reflect the limitations of perturbation
theory and indicate the presence of non--perturbative $\frac{1}{Q^2}$--power
corrections, as indeed the OPE in the physical vacuum suggests. What we are
proposing here is another way to approach the non--perturbative regime of QCD.
We want to explore the kind of ambiguities which
appear when the external $Q^2$ is chosen 
to be below $\Lambda_{L}^2$. Then, with
$Q^2 < \Lambda_{L}^2$, there is always a region in the virtual
$Q^2 \le k_{E}^2\le\infty$ integration region  which is also below
$\Lambda_{L}^2$, (the darker triangle in Fig.~8.) It is precisely this
integration region which is responsible for the promotion of the UV renormalons
to poles on the right hand side of the Borel plane. From considering IR
renormalons we are learning what type of corrections  to the perturbative
evaluation of two--point functions to expect, coming from the large $Q^2$
asymptotic regime. In the case of UV renormalons we are learning what type
of effective local operators govern the low--energy physics of the same
two--point function, when the virtual Euclidean momenta
$k_{E}^2$ is integrated all the 
way from $\infty$ down to the low--$Q^2$ regime .
The perturbative approach to these two {\it non--perturbative} limits is very
different, and it is therefore not surprising that the physics in the two cases
is also different. We insist on the fact that {\it both} limits involve an
integration region of virtual momenta {\it below} the Landau scale; and in that
respect there is no reason to expect one limit to be more physical than the
other.         

{}From another point of view let us notice that eq.~(\ref{eq:f}) implies that
$\alphaq \rightarrow 0^-$ as $Q^2\rightarrow 0$. This means that the
behaviour
of $\alphaq $ extracted by analytic continuation in the  low--energy region is
governed by the trivial IR  fixed point at $\alphaq =0^-$. In this sense it is
pretty much  like an abelian theory, only that the expansion parameter is 
$\mid\alphaq\mid = - \alphaq > 0$ ; 
(see eq. (\ref{eq:g}))~\footnote{The one--loop 
coupling $b_0 \alpha(Q^2) = \left(\log
Q^2/\Lambda_L^2\right)^{-1}$ is invariant under the change $\alpha(Q^2)
\rightarrow - \alpha(Q^2)$ , $b_0\rightarrow -b_0$.}. We know
that an abelian theory is not asymptotically free and consequently that its UV
renormalons sit on the right hand side of the Borel plane. This is what
happens here as well. It is then reasonable that, as seen from the extreme
low--energy end, there may be local higher dimensional operators suppressed by
the large scale $\Lambda_L$ since it is the only scale available. As the
energy goes up and eventually goes over this scale
$\Lambda_L$, the dynamics is ultimately governed by the true UV fixed point
of QCD at $\alphaq=0^+$, as follows from its nonabelian character. As far as
renormalons are concerned, if we had
to exemplify this, a similar situation could be like having QED embedded in a
nonabelian GUT model like, e.g., $SU(5)$. 


\vspace*{0.5cm}

We turn next to the discussion of the pseudoscalar two-point function in the
same low--$Q^2$ regime. The analysis runs parallel to the vector one. As far as
the two--chain renormalon is concerned one sees that the leading result
eq.~(\ref{eq:Ptwochainsk}) is proportional to the one in the vector channel,
eq.~(\ref{eq:eighteen}). One therefore concludes that the ambiguity is 
\be
\label{eq:bb}
\delta P(Q^2)\mid_{Q^2<k_{E}^2}^{\mbox{\rm \small two\ chains}} = - \cK \ 
\frac{N_c}{16\pi^2}\
\left(\frac{N_c}{2}\,\frac{1}{2\pi b_0}\right)^2 \
\frac{Q^2}{\Lambda_L^2}\ 
\left( 8 \log^{2}\frac{\Lambda_L^2}{Q^2} +
\cO \left(\log\frac{\Lambda^2_{L}}{Q^2}\right)\right)
\ee 
with $\cK$ the same constant appearing in eq.~(\ref{eq:b}). This ambiguity
can also be reproduced by a {\it local} four--fermion operator; an operator
exactly of the form of the scalar--pseudoscalar four--fermion operator which
appears in the ENJL model:
\be
\label{eq:cc}
\cO = \frac{8\pi^2 G_S}{N_c \Lambda_{\chi}^2}\ 
\sum_{\mbox{\rm \small a,b=flavour}} 
\left(\bar q^a_R(x) q^b_L(x) \right)\ \left(\bar q^b_L(x) q^a_R(x) \right) \
{}.
\ee 
The identification with the resulting perturbative ambiguity requires now 
\be 
G_S \equiv - \ 4 \cK \left(\frac{1}{2\pi b_{0}}\frac{N_c}{2}\right)^2\ .
\ee 

Assuming that the ambiguity $\cK$ is a measure of the final non--perturbative
value for the coefficients of the ENJL--like four--fermion operators would lead
to the conclusion that $G_S/G_V = 16$. However, one should not forget that this
result has been obtained at the two--chain renormalon level. As we shall next
show, higher number of chains also contribute with the same
$Q^2/\Lambda_L^2$ dependence as in eqs.~(\ref{eq:b},\ref{eq:bb}), with the
result that not even the ratio $G_S/G_V$ can be fixed from the lowest
non--trivial two--chain calculation .

When we examined UV renormalons in the high--$Q^2$ regime in
\ref{subsubsec:Comments} we saw  
that diagrams with an increasing number
of chains are not suppressed with respect to the contribution coming from two
chains. This feature persists when
$Q^2<
\Lambda_L^2$. Let us consider the integral in (\ref{eq:manychains}) again, but
now for $Q^2<
\Lambda_L^2$:
\be 
I_{p}= N_c Q^2\int_{Q^2}^{\infty}\ \frac{dk_E^2}{k_E^4}\ \left[N_c
\alphak\right]^p\ \log^2\frac{k_E^2}{Q^2}\ ,
\ee 
where we have 
introduced convenient factors of $N_c$ to match the dependence of  a leading
contribution at large $N_c$. Defining $w=-\alphaq b_0
\log(k_E^2/Q^2)$, where $\alphaq<0$ is given by equation (\ref{eq:f}), one
finds
\be 
I_{p}= N_c\ \frac{\left[ N_c \alphaq\right]^p}{\left[-b_0\alphaq  \right]^3} 
\int_0^{\infty}dw\  e^{-\frac{w}{\left(-b_0\alphaq\right)}}\
\frac{w^{2}}{\left(1-w\right)^p} \ .
\ee 
This integral has a pole at $w=1$ of multiplicity $p$. Therefore the
ambiguity is given by
\be
\delta I_{p} = \tilde 
\cK\ N_c\ \frac{\left[ N_c \alphaq\right]^p}{\left[-b_0\alphaq
\right]^3}\ \ \left(
\frac{d}{dw}\right)^{p-1}\left[e^{-\frac{w}{\left(-b_0\alphaq\right)}} w^{2}
\right]_{w=1}\ .
\ee
Each time the derivative $d/dw$ acts on the exponential it brings down 
a factor $(b_0\alphaq)^{-1}$, hence a power of $\log(Q^{2}/\Lambda_L^2)$. When
$Q^2\rightarrow 0$, the leading contribution is given by
\be
\label{eq:uncertainty}
\delta I_{p} = \frac{-1}{(p-1)!} \ 
\tilde \cK\ N_c\ \left(\frac{N_c}{b_0}\right)^{p}\
\frac{Q^2}{\Lambda_L^2}\ \left(\log^2\frac{\Lambda_L^2}{Q^2} +\cdots\right)\ ,
\ee where $\cdots$ stands for subleading terms when $Q^2<<\Lambda_L^2$.
Consequently one obtains once more the $Q^2$ dependence of eqs. (\ref{eq:b})
and (\ref{eq:bb})~\footnote{This persists even after use of a two--loop $\beta$
function. See Appendix A.}, and 
therefore the ambiguity can be reproduced by the
insertion of the ENJL operators (\ref{eq:c}) and (\ref{eq:cc}),  although with
different coefficients $G_S,G_V$. The calculation of these coefficients for an
infinite number of chains looks like a much harder problem since there is no
obvious parameter that dictates that a higher number of chains should give a
smaller contribution. Therefore there is no reason to believe the relation
$G_S=16 G_V$ will be preserved beyond two chains and, for instance, the $G_S$
and $G_V$ corresponding to the three--chain double--box diagram come in the
ratio $G_S/G_V = 64$.  From a practical point of view, one may just as well
leave   the coefficients $G_S$ and $G_V$ as free phenomenological parameters
to be fitted to some   experimental 
input \cite{BBdR,Bij}. This is just the usual
situation.  Renormalons help one to make good guesses for parameterizing
non--perturbative physics, but they are not capable of producing 
quantitative results: in our case, $G_S$ and
$G_V$ appear to be incalculable quantities. 

There is another point of view, 
concerning the ratio  $G_S/G_V$,  which has been
advocated by the authors of ref.~\cite{YZ94}. 
Their claim is that the consistency
of the successes of the QCD sum rules with 
the ENJL model requires $G_S\gg G_V$.
If that was the case, one may then perhaps 
be tempted to take more seriously the
quantitative perturbation theory results above. The analysis in
ref.~\cite{BBdR}, however, shows that the success of the ENJL model in
predicting the $\cO (p^4)$ $L_{i}$ coupling constants of the chiral Lagrangian
hinges on the fact that $G_V$ has to be as large as $G_S$~\footnote{It is in
particular the constants $L_5$ and $L_8$ what requires the axial coupling
constant  of the constituent quarks $g_{A}$ to be $g_{A}\simeq 0.6$ and hence
$G_{V}\sim G_{S}$.}. A detailed phenomenological 
analysis of this issue is under
investigation~\cite{PPdeR97}.

\newpage
\setcounter{equation}{0}

\section{Conclusions and Outlook.}
\label{sec:ConOut}

The analysis of ultraviolet renormalons, in the way we just described, seems
helpful in bridging the gap between high energy and low energy in QCD.
It offers an arena where it begins to be possible to discuss generic dynamical
patterns of the effective low energy chiral Lagrangian of QCD. At high energies
UV renormalons hint at the emergence of dimension $d=6$ four--fermion operators
{}from the large orders of perturbation theory~\cite{VZ94a,VZ94b}. What we have
seen here is that, at low energies, they may even ``feel" the existence of the
scale
$\Lambda_{\chi}$ in the normalization of these four--fermion operators. 
There is a correspondence between the leading $n!$ behaviour at large orders of
perturbation theory and the leading $Q^2\rightarrow 0$ 
behaviour as both are due
to the insertion of the same effective operator.

Perturbation theory (to all orders) and 
analyticity in the coupling constant has led us this far.  Let us
now turn to some physical comments. The picture which emerges is obviously
incomplete as, so far, there is no signal of spontaneous  chiral symmetry
breaking or confinement. Neither of these features should be expected
however from
perturbation theory arguments alone, which after all is the framework of
renormalon calculus. As we have seen, the analysis of UV renormalons in the
regime of {\it low external $Q^2$ momenta} provides a plausible link with the
ENJL model. This model however is only supposed  to be meaningful in the event
of a clear separation between the confinement scale
$\Lambda_{\mbox{\rm conf}}$ and the spontaneous  
chiral symmetry scale $\Lambda_{\chi}$,
as it is an effective Lagrangian description in between the two scales with 
 $\Lambda_{\mbox{\rm conf}} << \Lambda_{\chi}$.  
That these two scales may be separated
widely enough seems to be backed up  by the phenomenological success of the
constituent chiral quark model\cite{GeM}, which follows naturally from the
ENJL model, and  wherein there is a succesful description in terms
of  ``constituent massive quarks''. For this to be possible within the UV
renormalon approach which has been advocated above, the coupling constant
$G_S$ in the
$d=6$ four--fermion operator (\ref{eq:cc}) has to be larger than unity. It is
very difficult to justify why $G_S$ should have this size within perturbation 
theory alone. However we know that
there is a (rather similar!) case ---the gluon condensate--- 
where the ambiguities
foreseen in perturbation theory via the IR renormalon analysis are finally
realized by a rather large vacuum expectation 
value of non--perturbative origin.
There also, renormalons only signal the appearance of non--perturbative
contributions in the form of
$Q^{-4}$ terms, and it is then a matter 
of non--perturbative dynamics what finally makes
the numerator large. It is not unreasonable to assume that the coupling
constants $G_S$ and $G_V$ of the four--fermion 
operators which govern the leading
UV renormalon effects may behave similarly to the gluon condensate and
turn out to have large values as well in the real world.


We have seen that, at least within the framework of the ``amputated'' effective
action in (\ref{eq:appr}), four--fermion operators govern the properties of the
leading UV renormalon. Assuming that, eventually beyond
perturbation theory, chiral symmetry is spontaneously broken by
the dimension--six scalar four--fermion operator with $G_S\ge 1$, the question
one may ask is:  can one safely neglect operators with dimensions higher than
six?  Although they are associated with subleading  renormalons in perturbation
theory; beyond perturbation theory the counting of dimensions changes for the
dimension--six scalar four--fermion operator with $G_S\ge 1$. Its effects
become suppressed   only by a
$1/\log(\Lambda^2_{\chi}/Q^2)^2$ instead of $Q^2/\Lambda_{\chi}^2$.
This enhancement partially takes place also in scalar four--fermion  operators
with higher dimensions. The net effect there, however, 
is that these  only modify
somewhat the relationship between the mass of the scalar particle and the mass
of the constituent quark  ($M_S\not= 2 M_{Q}$) but they do not change the
spectrum or the interactions\cite{H}. In the vector channel, higher--than--six
dimensional operators only modify the results of the ENJL model at values of
$Q^2\sim M_{\rho}^2$\cite{PP} but not at lower energy.    

More work than what we have presented here is clearly needed and seems
worthwhile. What is at stake is that it may be finally  possible to understand
why the ENJL model, and hence the 
constituent chiral quark model, does so well in
describing low--energy QCD phenomenology~\cite{BBdR,Bij}. We think that the
patterns unraveled so far must not be just a coincidence.

\vspace*{2cm}

\begin{center}
{\Large {\bf Acknowledgements}}
\end{center}

\vspace{10pt}

One of us (S.P.) would like to thank P.~Hasenfratz for reminding him of the
relevance (in the ordinary sense of the word) 
of four--fermion operators in the case of QED, V.~Azcoiti for updating
him on these matters and D.~Espriu for conversations.
We also thank J.~Bijnens, A.~Pich and J.~Watson 
for discussions at various stages of this work. 
Part of this work was done at the Benasque Center for Physics, and
we wish to acknowledge the very stimulating atmosphere there which helped us
much.  E.~de R. has benefited from the ``de Betancourt--Perronet'' prize for
visits to Spain.
        
\newpage
\vspace{20pt}


\appendix
\setcounter{equation}{0}
\def\theequation{A.\arabic{equation}}

{\Large {\bf Appendix A}}

\vspace{10pt}

\noi
In this appendix we shall outline how to generalize our previous results to
the case of a two--coefficient $\beta$ function such as
\be
\label{eq:betatwo}
\alpham\beta(\alpham) = - b_0 \alpham^2 - b_1 \alpham^3 \ .
\ee The use of this  $\beta$ function is interesting since i) it has a finite
radius of convergence in $\alpha$ \cite{Beneke93} (it is a polynomial), and
ii) any $\beta$ function can be brought to this form by making a perturbative
redefinition of the coupling constant \cite{tH78}.

The equation
\be
\frac{d\alpham}{d\log(\mu^2/Q^2)} = - b_0 \alpham^2 - b_1 \alpham^3
\ee can be integrated to yield
\be
\label{eq:RGE}
\frac{1}{b_0} \left( \frac{1}{\alpham} - \frac{1}{\alphaq}\right) = 
\log(\mu^2/Q^2) + \frac{b_1}{b_0^2} \log \frac{\frac{1}{\alpham} +
\frac{b_1}{b_0}}{\frac{1}{\alphaq} + \frac{b_1}{b_0}}\ .
\ee This equation has a Landau pole at 
\be
\Lambda_L^2= \mu^2\ e^{- \frac{1}{b_0 \alpham}} \ 
\left(1+ \frac{b_0}{b_1\alpham}\right)^{\frac{b_1}{b_0^2}}\ .
\ee

Notice that equation (\ref{eq:RGE}) implies that
\be
\alpham \simeq \frac{1}{b_0 \log(\mu^2/\Lambda_L^2)}\ ,
\ee both when $\mu^2\rightarrow + \infty$ and $\mu^2\rightarrow 0$.

The integral we want to study is 

\be I= \int_{Q^2}^{\infty}\ \frac{dk_E^2}{k_E^2}\ \left(
\frac{k_E^2}{Q^2}\right)^n\ 
\left[\alpha(k_E^2)\right]^p\ \log^2\frac{k_E^2}{Q^2}\ ,         
\ee         for $n$ a negative integer with $n\leq -1$ and $p$ a natural
number, 
$p\geq 2$.

Making the change of variables \cite{Gr96} 
\be
\frac{z}{z_n}\equiv \frac{\frac{1}{\alphaq} -
\frac{1}{\alpha(k_E^2)}}{\frac{1}{\alphaq} - \frac{b_1}{b_0}}\ ,
\ee where $z_n\equiv n/b_0 < 0$ and $Q^2 << \Lambda_L^2$, so $\alphaq \simeq 
(b_0 \log(Q^2/\Lambda_L^2))^{-1}$, one obtains
\bea
\label{eq:integral} I = \int_0^{-\infty}&& \frac{dz}{-n}\ \frac{e^{-z\left(
\frac{1}{\alphaq} + \frac{b_1}{b_0}\right)} }{\left( 1 - \frac{z}{z_n}
\right)^{1+\delta_n}}\nn \\ & & \left[\frac{1}{\alphaq}-\frac{z}{z_n}\left(
\frac{1}{\alphaq}+\frac{b_1}{b_0}\right)\right]^{1-p}\nn \\ & & \qquad \left[-
\frac{z}{z_nb_0\alphaq} -
\frac{zb_1}{z_nb_0^2}-\frac{b_1}{b_0^2}
\log\left( 1- \frac{z}{z_n}\right)\right]^{2}
\eea where $\delta_n\equiv nb_1/b_0^2<0$.

Since $z_n<0$, the integration region contains two singularities that, in the
limit $\alphaq \rightarrow 0^-$, collapse to a branch point at $z=z_n$.
Therefore
\bea
\delta I =&&\alphaq^{p-1}\int_0^{-\infty} \frac{dz}{-n}\nn \\
 & &\mbox{\rm Disc}\left\{ \frac{e^{-z\left(
\frac{1}{\alphaq} + \frac{b_1}{b_0}\right)} }{\left( 1 - \frac{z}{z_n}
\right)^{p+\delta_n}}\left[-\ \frac{z}{z_nb_0\alphaq} -
\frac{zb_1}{z_nb_0^2}-\frac{b_1}{b_0^2}
\log\left( 1- \frac{z}{z_n}\right)\right]^{2}\right\}
\eea hence
\be
\delta I \simeq \alphaq^{p-1}\int_0^{-\infty} \frac{dz}{-n}  \mbox{\rm
Disc}\left\{
\frac{e^{-z\left(
\frac{1}{\alphaq} + \frac{b_1}{b_0}\right)} }{\left( 1 - \frac{z}{z_n}
\right)^{p+\delta_n}}\left[-\ \frac{z}{z_nb_0\alphaq}\right]^{2}\right\}
\ee
which can still be approximated as
\be
\delta I \simeq \ \frac{[\alphaq]^{p-3}}{b_0^{2}}
\int_0^{-\infty} \frac{dz}{-n}
 e^{-z\left(\frac{1}{\alphaq} + \frac{b_1}{b_0}\right)}  
\mbox{\rm Disc} \left\{
\left( 1 - \frac{z}{z_n}\right)^{-p-\delta_n}\right\}\ .
\ee                                         
Given that $\mbox{\rm Disc} \left(
1 -
\frac{z}{z_n}\right)^{-p-\delta_n} = \hat
\cK (1-e^{-i2\pi\delta_n}) \mid 1-z/z_n\mid ^{-p-\delta_n}$, where $\hat 
\cK$ is an arbitrary constant, one obtains
\be
\label{eq:result}
\delta I \simeq  \ \frac{\hat \cK (1-e^{-i2\pi\delta_n})}{-n 
(b_1/b_0)^{\delta_n}}\ z_n^{p+\delta_n}\ e^{-z_nb_1/b_0}\
\Gamma(1-p-\delta_n)\ \left(\frac{Q^2}{\Lambda_L^2}\right)^{-n}\
\log^2\frac{\Lambda_L^2}{Q^2}\ + \cdots\,.
\ee where the ellipses stand for terms subleading in the limit
$Q^2/\Lambda_L^2\rightarrow 0$. One can check that the $b_1\rightarrow 0$
limit of the previous expression is equivalent to eq.~(\ref{eq:uncertainty})
in the text, which was obtained in the case $b_1=0$.

One finally sees from eq.~(\ref{eq:result}) that the leading contribution
($n=-1$) produces again an ambiguity $\sim (Q^2/\Lambda_L^2)
\log^2(\Lambda_L^2/Q^2)$, exactly as in the $b_1=0$ case discussed in the
text. Therefore our conclusions are also valid in the general case of the
$\beta$ function (\ref{eq:betatwo}).  

\vspace{20pt}

\appendix
\setcounter{equation}{0}
\def\theequation{B.\arabic{equation}}

{\Large {\bf Appendix B}}

\vspace{10pt}

\noi
We would like to comment here on the issue of the ``freezing" of the strong
coupling constant at low energy.

Sometimes one finds in the literature~\cite{MaS94} proposals for a 
non--perturbative behaviour of the 
coupling constant in the low--energy region of
the form
\be
\label{eq:freezing}
\tilde \alphaq = \frac{1}{b_0 \log \frac{Q^2 + C^2}{\Lambda^2}}\ .
\ee This form clearly guarantees the right high-$Q^2$ behaviour dictated by
perturbation theory. It can be obtained from the following RG equation:
\be
\label{eq:betafreezing} Q^2\ \frac{d\tilde\alphaq}{dQ^2} =\tilde{\alphaq} \beta
\Big(\tilde
\alphaq\Big)\ .
\ee with 
\be
\beta \Big(\tilde \alphaq\Big) = \ -\ b_0 \tilde \alphaq 
\left[1-\frac{C^2}{\Lambda^2} e^{- \frac{1}{b_0\tilde \alphaq}}\right]\ .
\ee The constant $C$ is supposed to encode the non--perturbative dynamics. As
$C^2 \rightarrow 0$ one recovers the usual situation in perturbation theory at
one loop. In this equation $C^2 > \Lambda^2 > 0$ so that $\tilde
\alphaq$ does not have a Landau pole.  

What we would like to point out here is that chiral symmetry breaking is a
strong constraint for this proposal, at least in the simplest form of
eq.~(\ref{eq:freezing}), as we now explain.

If we take the Adler function $\cA(Q^2)$ (defined in the main text) one knows
{}from general arguments of spontaneous chiral symmetry breaking in the
large--$N_c$ limit  that $\cA(Q^2) \sim Q^2$ as $Q^2\rightarrow 0$ \footnote{In
the large-$N_c$ limit only one--meson states contribute to $\cA(Q^2)$. Since
these states are not massless in the chiral limit, dimensional analysis imposes
that 
$\cA(Q^2)
\sim Q^2/M^2$ with $M$ the corresponding  meson mass.}, and in particular
$\cA(0)=0$. 

In perturbation theory (i.e. at large $Q^2$) the Adler function is given  by
\be
\label{eq:adlerapp}
\cA(Q^2) = \frac{1}{8\pi^2}\ \left(1 + \frac{\alphaq}{\pi}\right)\ ,
\ee where $\alphaq$ is the usual perturbative coupling constant at one loop,
say. If $\tilde \alphaq$ is supposed to incorporate all the non--perturbative
dynamics by deviating from $\alphaq$ as $Q^2\rightarrow 0$ one would expect
that
\be
\label{eq:adlerappprime}
\cA(Q^2) = \frac{1}{8\pi^2}\ \left(1 + \frac{\tilde \alphaq}{\pi}\right)\ ,
\ee would be the natural answer. However we saw that freezing leads   to
$\tilde \alphaq \rightarrow $ constant ($>0$) \footnote{As a matter of fact
$\tilde \alpha(0)/\pi \lesssim 0.3$ \cite{MaS94}; i.e. a relatively small
correction to unity.}as $Q^2\rightarrow 0$ and
therefore $\cA(0) \not= 0$, in conflict with chiral symmetry.

The problem seems rather generic for coupling constants obeying an equation
such as (\ref{eq:betafreezing}) for if  $\cA(0)$ has to vanish, 
$\tilde\alphaq$ has to become negative at low $Q^2$.  But since $\tilde
\alphaq$ is positive at high $Q^2$ it follows, if $\tilde \alphaq$ is a
continuous function of its variable, that there must be an intermediate point
at which 
$\tilde \alphaq$ is small and its slope is positive, which is impossible since
its slope is the $\beta$ function which, for small $\tilde \alphaq$, has to be
negative according to asymptotic freedom.

We would like to emphasize that the argument we presented above is not
intended as a  ``NO-GO" theorem but as a sort of illustrative warning signal.
One way out is to invalidate eq.~(\ref{eq:adlerappprime}), because more  powers
of $\tilde \alphaq$ cannot be neglected or for any other reason.   For
instance, there are models that are successful in describing diffractive
phenomena  in which it is argued that the freezing coupling  constant
(\ref{eq:freezing}) also comes with a $Q^2$-dependent dynamical mass for the
gluon. In this case eq.~(\ref{eq:adlerappprime}) does not  hold and one may
then be safe. As it turns out, however, in this case  it takes a more detailed
analysis of the Schwinger--Dyson equations to reveal that, actually,
these models do not seem capable of generating enough chiral symmetry
breaking in the end~\cite{NR96}.

Another possibility is that the freezing of the coupling constant may have a
more sophisticated dynamical origin. For instance, if one allows that the 
RG equation be governed by a $\beta$ function that depends on {\it both}
$\tilde \alphaq$ and $Q^2$, $\beta\Big(\tilde\alphaq , Q^2\Big)$, then 
eq.~(\ref{eq:adlerappprime}) may as a matter of fact reproduce $\cA(0)=0$. Just
as an  existing proof, let us quote one such
$\tilde \alphaq$:
\be
\tilde \alphaq = - \pi e^{-\frac{Q^2}{\Lambda^2}} +
\frac{e^{-\frac{\Lambda^2}{Q^2}}}{b_0 \log\left(a +
\frac{Q^2}{\Lambda^2}\right)}
\ee for arbitrary parameters $\Lambda^2$ and $a>1$. 

Therefore our conclusion is that spontaneous chiral symmetry breaking imposes
rather severe restrictions on the idea of freezing. One should check
that enough
symmetry breaking can be produced before any argument relying on a
particular freezing coupling constant is put forward.

\newpage


\end{document}